\documentclass[conference]{IEEEtran}

\usepackage{mathptmx}        % Times font
\usepackage{amsmath, amssymb}
\usepackage{physics}
\usepackage{xcolor}
\usepackage{tikz}
\usepackage{booktabs, makecell, multirow, diagbox}
\usepackage[linesnumbered,ruled,vlined]{algorithm2e}
\usepackage[noend]{algpseudocode}

\usepackage[normalem]{ulem}  % For underlining
\usepackage[final]{microtype}
\usepackage{fancyhdr}
\usepackage[keeplastbox]{flushend}
\usepackage{titlesec}
\titleformat{\title}{\normalfont\LARGE\bfseries}{}{}{}

\usepackage[hyphens]{url}
\usepackage[sort,nocompress]{cite}
\usepackage[bookmarks=true, breaklinks=true, letterpaper=true,
            colorlinks, citecolor=blue, linkcolor=blue, urlcolor=blue]{hyperref}
\usepackage[capitalise]{cleveref}

\newcommand{\frameworkname}{iSwitch}
\newcommand{\NISQname}{NISQ-Bare}
\newcommand{\MSDname}{MSD-Logical}
\newcommand{\INJname}{INJ-Logical}
\newcommand{\BLname}{CodeSwitch}
\newcommand{\Movename}{LogicMove}
\newcommand{\CNOTname}{LogicCX}

\title{\frameworkname: QEC on Demand via In-Situ Encoding of Bare Qubits for Ion Trap Architectures}

\author{
\IEEEauthorblockN{
Keyi Yin\IEEEauthorrefmark{1},
Xiang Fang\IEEEauthorrefmark{1},
Zhuo Chen\IEEEauthorrefmark{1},
David Hayes\IEEEauthorrefmark{2},
Eneet Kaur\IEEEauthorrefmark{3},
Reza Nejabati\IEEEauthorrefmark{3}, 
Hartmut Haeffner\IEEEauthorrefmark{4}, \\
Wes Campbell\IEEEauthorrefmark{5},
Eric Hudson\IEEEauthorrefmark{5},
Jens Palsberg\IEEEauthorrefmark{5},
Travis Humble\IEEEauthorrefmark{6},
Yufei Ding\IEEEauthorrefmark{1} \\ \\
}

\IEEEauthorblockA{\IEEEauthorrefmark{1}University of California, San Diego, CA, USA}
\IEEEauthorblockA{\IEEEauthorrefmark{2}Quantinuum, Broomfield, CO, USA}
\IEEEauthorblockA{\IEEEauthorrefmark{3}Cisco Quantum Lab, San Jose, CA, USA}
\IEEEauthorblockA{\IEEEauthorrefmark{4}University of California, Berkeley, CA, USA}
\IEEEauthorblockA{\IEEEauthorrefmark{5}University of California, Los Angeles, CA, USA}
\IEEEauthorblockA{\IEEEauthorrefmark{6}Oak Ridge National Laboratory, Oak Ridge, TN, USA}
}

\begin{document}
\maketitle
\pagestyle{plain}

% ---------- Main Sections ----------
\begin{abstract}
Recent advances in quantum hardware and error correction have paved the way for early fault-tolerant (EFT) quantum computing. We propose \frameworkname, a hybrid system architecture for trapped-ion quantum computers (TIQC) that exploits ultra-high-fidelity single-qubit gates and efficient logical CNOTs enabled by ion shuttling. \frameworkname~employs bare qubits for single-qubit operations and QEC-encoded logical qubits for two-qubit gates, avoiding full logical encoding, gate synthesis, and magic state distillation. To enable this selective encoding, we develop a low-noise conversion protocol between bare and logical qubits, a hybrid instruction set tailored to two-dimensional (2D) TIQC layouts, and a compiler that minimizes conversion overhead and optimizes scheduling. 
Evaluations on variational quantum algorithm benchmarks show that \frameworkname~achieves comparable fidelity to conventional QEC methods, while reducing qubit and operation counts by roughly 33–50\%, offering a practical, resource-efficient path toward EFT quantum computing.

% Recent advances in quantum hardware and quantum error correction (QEC) have set the stage for early demonstrations of fault-tolerant quantum computing (FTQC). A key near-term goal is to build a system capable of executing millions of logical operations reliably—referred to as a megaquop quantum computer (MQC). In this work, we propose a novel system architecture targeting MQC on trapped-ion quantum computers (TIQC), leveraging their ultra-high-fidelity single-qubit gates (1Q) and efficient two-qubit (2Q) logical CNOT gates enabled by the quantum charge-coupled device (QCCD) architecture with the ion shuttling feature.

% We propose~\frameworkname, a hybrid encoding scheme that uses bare qubits for 1Q gates and QEC-encoded logical qubits for 2Q gates. 
% This approach avoids fully encoding all qubits, eliminating the overhead of gate synthesis, teleportation, and magic state distillation for non-Clifford gates. To support this, we design (1) a low-noise conversion protocol between bare and logical qubits, (2) a bare-logical hybrid instruction set architecture tailored for 2D grid-based TIQC, and (3) a compiler that minimizes conversion cost and optimizes the scheduling efficiency. We evaluate our approach on VQA and small-scale FTQC benchmarks, showing that it achieves superior performance improvements with significantly reduced resource overhead, offering a practical path toward early FTQC on TIQC.

% \textcolor{red}{(Jens: the title talks about ``Encoding of Bare Qubits'' but the previous sentence talks about ``unencoded bare qubits''; seems different.)} 
\end{abstract}
\section{Introduction} \label{sec: intro}
Quantum noise remains a fundamental barrier for today’s quantum computing \cite{arute2019quantum, stilck2021limitations, de2023limitations}. 
Various noise channels accumulate across circuit operations~\cite{google2023suppressing, acharya2024quantumerrorcorrectionsurface}, making the execution of large-scale quantum programs still practically infeasible, although modern quantum devices now already host hundreds to thousands of physical qubits across diverse hardware platforms~\cite{arute2019quantum, huang2020superconducting, bravyi2022future, bruzewicz2019trapped, chen2023benchmarking, wurtz2023aquila}. Programs that leave more qubits unprotected are limited to shallower circuit depths~\cite{temme2017error, endo2018practical}. This accumulation constrains the practical benefits of larger devices, showing that simply increasing qubit count does not extend computational capability.

Quantum error correction (QEC)~\cite{calderbank1996good, bombin2006topological, kitaev2003fault, fowler2012surface, bravyi2024high} is widely regarded as the only viable path to overcome this bottleneck and enable fault-tolerant quantum computing (FTQC)~\cite{nielsen2010quantum, gottesman1998theory, shor1996fault, koukoulekidis2023framework}, which is essential for solving large-scale, practically useful problems~\cite{shor1999polynomial, grover1996fast, cao2019quantum, daley2022practical}. With sufficient resources to implement QEC, a fault-tolerant quantum computer can in principle execute arbitrarily long circuits while maintaining high overall fidelity~\cite{google2023suppressing, gidney2021factor}.

However, current fault-tolerant schemes, which require every operation to be executed fault-tolerantly, are extremely resource-intensive \cite{google2023suppressing, gidney2021factor}. The resource overhead primarily arises from two sources. First, each logical qubit must be encoded into a sufficient number of physical qubits to ensure its protection during execution. In particular, commonly used surface code encodings~\cite{fowler2012surface} are highly resource-intensive, requiring many physical qubits per logical qubit to achieve practical error rates.
Second, the Eastin-Knill theorem~\cite{eastin2009restrictions} implies that additional resources are necessary to implement gates not native to the QEC code, such as the non-Clifford $T$ gate in the surface code. The standard approach for realizing these non-Clifford operations, \emph{magic state distillation} (MSD) \cite{bravyi2005universal, bravyi2012magic, haah2018codes}, consumes a large number of ancilla qubits and significant runtime \cite{litinski2019magic}. Overall, a fully FT design may require on the order of millions of physical qubits.

\begin{figure}[!ht]
\centering
\includegraphics[width=0.47\textwidth]{Fig/Overview.pdf}
\caption{Overview of~\frameworkname.}
\label{fig:arch_overview}
\end{figure}

Unsurprisingly, the resource demands of full FT schemes far exceed near-term device capabilities. As noted in a recent talk by John Preskill~\cite{preskill2025beyond}, the next generation of quantum machines, dubbed Megaquop Quantum Computers (MQCs), is expected to host thousands of physical qubits, offering far greater scale than current noisy intermediate-scale quantum (NISQ) devices~\cite{preskill2018quantum} while still falling short of full fault-tolerant capability. This intermediate regime motivates early fault-tolerant (EFT) strategies~\cite{dangwal2025variational, katabarwa2024early,lao2022magic}, which aim to move beyond NISQ and enable practical quantum applications by incorporating some QEC.

Among these EFT strategies, \emph{state injection}~\cite{dangwal2025variational,lao2022magic, toshio2025practical} has emerged as a particularly promising approach. By implementing non-Clifford operations such as $R_z$ gates at reduced fidelity, state injection avoids costly magic state distillation factories, freeing resources for logical qubit encoding. This deliberate trade-off between performance and resource consumption allows EFT schemes to enhance algorithmic fidelity on near-term devices while circumventing the prohibitive overhead of full fault tolerance. Rather than providing a universal optimization for arbitrary circuits, state injection specifically targets applications with sparse non-Clifford gates, such as VQE and QAOA, promising candidates for demonstrating near-term quantum advantage~\cite{cerezo2021variational, peruzzo2014variational, farhi2014quantum}.

% Rather than providing a universal optimization for arbitrary circuits the state injection method targets a practically important class of workloads in which non-Clifford gates are sparse and CNOT gates dominate. Instead of allocating substantial resources to magic state distillation for a few $R_z$ gates, the method prioritizes fault tolerance for the more frequent CNOT operations by using higher code distances. 

However, the state injection approach remains practically limited. It still incurs additional resource overhead, including QEC resources needed to protect the ancilla logical qubit during state preparation, as well as extra quantum operations and routing required for the injection circuit~\cite{dangwal2025variational}. Moreover, it is a general method that does not fully exploit current hardware characteristics, such as native gate fidelities, connectivity, and noise profiles~\cite{acharya2024quantumerrorcorrectionsurface, google2023suppressing, zhao2022realization, Bluvstein2024}.

These limitations motivate the development of more efficient, hardware-informed EFT schemes, in which QEC strategies are tailored to leverage the native strengths of near-term quantum hardware. To illustrate this approach, we focus on a leading platform, trapped-ion quantum computers (TIQC) \cite{bruzewicz2019trapped, strohm2024ion, moses2023race}, which offer system-level features that can be directly exploited to reduce EFT overhead. In particular, we highlight three key observations:
(1) TIQC’s ultra-accurate single-qubit gates already meet EFT fidelity targets (99.9999\%~\cite{harty2014high}), allowing non-Clifford gates to run on bare qubits and avoiding the overhead of magic state distillation and state injection.  
(2) TIQC’s grid-based QCCD (quantum charge-coupled device) architecture aligns naturally with the 2D layout of surface code patches, supporting efficient logical CNOT gates. Its ion-shuttling capability enables transversal operations by bringing two surface code patches together and applying physical CNOT gates between corresponding qubits, significantly reducing the cost of multi-qubit logical operations.  
(3) Surface codes support dynamic encoding transitions. Existing code deformation protocols have demonstrated the possibility of transitioning between bare and logical encodings~\cite{lao2022magic, li2015magic, choi2023fault, gidney2024magic, vuillot2019code, yin2024surf}. Although these protocols are too costly for direct use, they inspire the development of runtime encoding-switch schemes compatible with on-the-fly execution.

Building on these insights, we propose \frameworkname, a hybrid QEC framework for the TIQC of QCCD architecture—the next-generation device from Quantinuum—that achieves EFT with minimal resource overhead for non-Clifford gates. \frameworkname~employs a \emph{temporal selective QEC} strategy.
This selective protection is applied both \textit{spatially}—encoding only qubits that require fault tolerance—and \textit{temporally}—allowing qubits to transition between bare and logical states according to circuit requirements. Together, these strategies achieve a balanced trade-off between fidelity and resource efficiency.
As shown in \cref{fig:arch_overview}, its design consists of three synergistic modules:
(1) \textbf{Runtime Encoding Protocol (Sec.~\ref{sec:encoding})} enables low-noise transitions between bare and logical qubits, allowing implementation of $R_z(\theta)$ gates without additional ancilla logical qubits.
(2) \textbf{Hardware-aware Hybrid ISA (Sec.~\ref{sec:hybrid-isa})} partitions the system into bare, logical, and interface regions, providing tailored instructions to coordinate operations and encoding transitions while abstracting hardware details, relieving the compiler from scheduling individual physical qubits.
(3) \textbf{Compiler with Dynamic Encoding (Sec.~\ref{sec:compiler})} manages encoding switches, qubit routing, and gate scheduling in a fidelity- and resource-aware way, minimizing noisy code-switch operations and optimizing space-time resources to maximize selective QEC performance.

% Building on these insights, we propose \frameworkname, a hybrid QEC framework for the TIQC of QCCD architecture, the next gerneration devicde of quantinuum, that achieves EFT with minimal resource overhead for non-Clifford gates. \frameworkname~employs a \emph{temporal selective QEC} strategy, executing single-qubit gates on bare qubits and two-qubit gates on encoded logical qubits. As shown in \cref{fig:arch_overview}, its design consists of three synergistic modules: (1) a \textbf{runtime encoding protocol} enabling low-noise conversion between bare and logical qubits, (2) a \textbf{hardware-aware hybrid ISA} partitioning the system into bare, logical, and interface regions for seamless hybrid execution, and (3) a \textbf{compiler} orchestrating encoding switches based on program demands and hardware constraints. Together, these modules enable practical hybrid QEC execution on TIQC, balancing fidelity and resource efficiency for near-term demonstrations of EFT.

In summary, this paper makes the following contributions:
\vspace{-3pt}
\begin{itemize}
\item We propose \textbf{\frameworkname}, a framework for TIQC that enables EFT by applying QEC selectively in space and time. \frameworkname~avoids the overhead of monolithic FTQC while satisfying EFT fidelity requirements.
\item We develop a runtime protocol for in-situ encoding conversion, a hybrid ISA for managing bare–logical execution, and a compiler that co-optimizes encoding allocation and routing, together enabling the selective QEC strategy of~\frameworkname.
\item Evaluation shows that~\frameworkname~achieves up to 43$\times$ performance improvement across various EFT applications compared to NISQ execution, while using 50\% and 33\% fewer qubits than standard FTQC and previous EFT methods, respectively, under the same qubit budget, demonstrating its practical utility.
\end{itemize}

\section{Background and Related Work}
This section begins by providing essential background on the value of EFT applications (Sec.~\ref{subsec: EFT}) and the fundamental bottlenecks of standard FT schemes (Sec.~\ref{subsec: FTQC}), focusing on surface codes. We then review the capabilities of TIQC hardware that make it especially suitable for near-term EFT (Sec.~\ref{subsec: TIQC}). 
Together, these elements set the foundation for our proposed solution of leveraging TIQC for EFT, which is presented in Sec.~\ref{sec:encoding}, \ref{sec:hybrid-isa}, \ref{sec:compiler}.

\subsection{EFT Applications}{\label{subsec: EFT}}
Early fault tolerance (EFT) has emerged as the next milestone on the roadmap to full-scale FTQC, targeting near-term systems capable of executing around million robust operations with QEC protection~\cite{preskill2025beyond}. This shift is driven by two key developments in recent years. First, the limitations of NISQ devices have become better understood~\cite{stilck2021limitations, de2023limitations}, reinforcing the view that quantum advantage is unlikely to be achieved under current noise levels without some form of fault tolerance. Second, a growing number of core FTQC building blocks have been successfully demonstrated across multiple platforms~\cite{acharya2024quantumerrorcorrectionsurface, Bluvstein2024, reichardt2024logical, postler2022demonstration, reichardt2024demonstration, kang2023quantum, rodriguez2024experimental}, suggesting that the field is now technically ready to transition from NISQ to EFT.

Despite its limited scale, EFT offers a promising path toward practical quantum advantage. Recent studies provide growing evidence that higher-accuracy VQAs, equipped with FT execution, can outperform classical heuristics by achieving better scaling and more accurate approximations to ground-state energies~\cite{he2024performance, shaydulin2024evidence, boulebnane2024solving, omanakuttan2025threshold, hao2024end, he2023alignment}. This is particularly relevant for applications in quantum chemistry~\cite{bauer2020quantum, pearson2020simulating, beverland2022assessing, vqe_qiskit_repo} and condensed matter physics~\cite{singh2020ising, bonechi1992heisenberg}, where classically intractable systems may become accessible through modest-scale quantum computations. Additionally, Hamiltonian simulation in areas such as quantum dynamics and quantum chemistry has demonstrated compelling accuracy improvements over classical approximation methods~\cite{bauer2020quantum, pearson2020simulating, beverland2022assessing}. These developments underscore the practical value of EFT and motivate system-level innovations specifically designed to support this emerging computational regime.

\subsection{Bottleneck of Standard FTQC Scheme}\label{subsec: FTQC}

This subsection focuses on surface codes to illustrate the substantial overheads of standard FTQC protocols, which presents a major barrier to practical implementation on near-term hardware. The dominant costs arise from the space overhead of logical encoding and the space-time overhead of non-Clifford logical T gate.

\begin{figure}[!ht]
\centering
\includegraphics[width=0.47\textwidth]{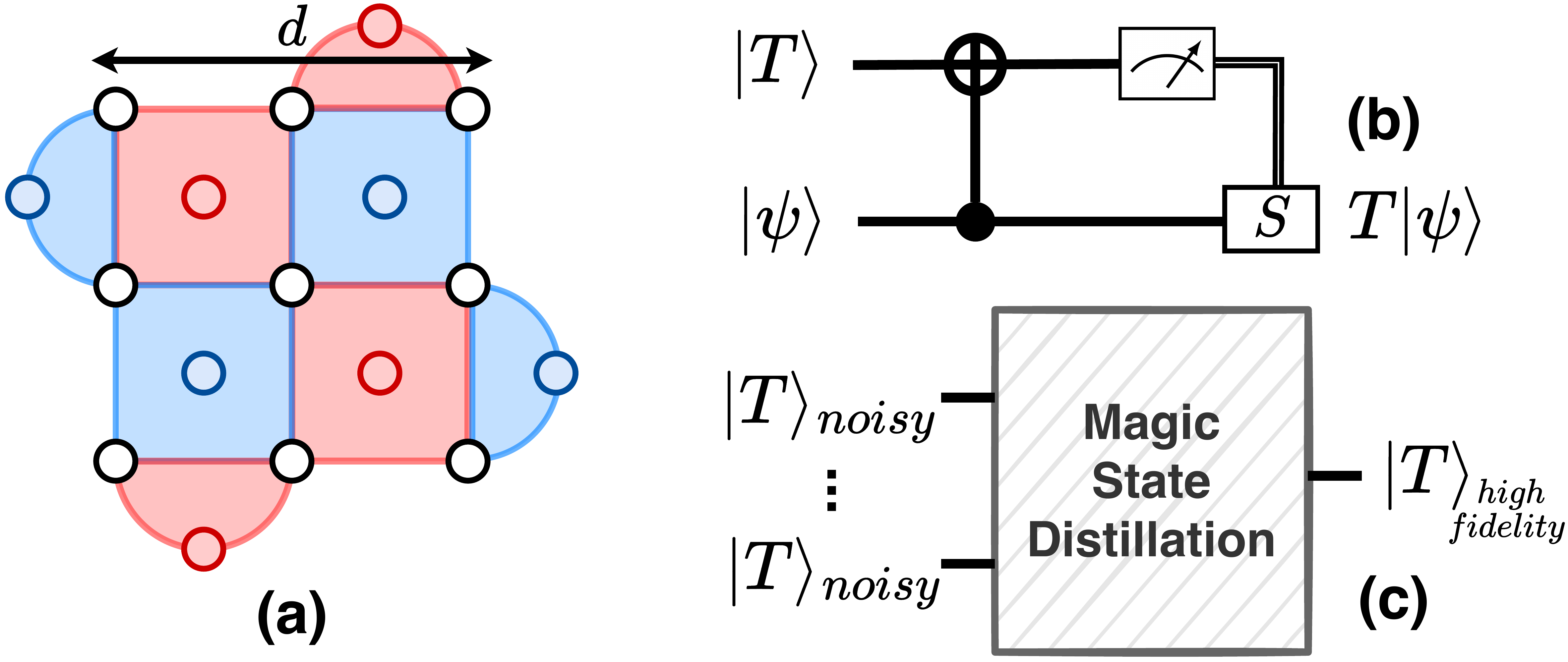}
\caption{(a) Surface code. (b) Gate Teleportation for T gate. (c) Magic state distillation.}
\label{fig: FTQCBack}
\end{figure}

\vspace{3pt}
\noindent\textbf{Logical Encoding.}
Surface codes are a leading candidate for FTQC due to their 2D local structure, efficient decoding algorithms~\cite{horsman2012surface, vuillot2019code, higgott2022pymatching}, and high threshold ($\sim 1\%$)~\cite{bravyi1998quantum, fowler2012surface}. 
However, they exhibit low encoding efficiency: a square patch of $2d^2-1$ physical qubits encodes only a single logical qubit, where $d$ is the code distance, and larger $d$ is required to achieve lower logical error rates (LERs). 
However, they exhibit low encoding efficiency (Fig.~\ref{fig: FTQCBack}(a)): a square patch of $2d^2-1$ physical qubits encodes only a single logical qubit, where $d$ is the code distance, and larger $d$ is required to achieve lower logical error rates (LERs). 
Simulations suggest that under a physical error rate of $10^{-3}$, reaching LERs of $10^{-4}, 10^{-5}$, and $10^{-6}$ requires distances $d=$5, 7, and 9, respectively~\cite{fowler2018low}. In practice, however, more complex noise structures demand even larger distances. For example, Google’s experiment~\cite{google2023suppressing} shows that achieving an LER of $\sim 10^{-4}$ requires a distance-13 surface code, corresponding to over 300 physical qubits per logical qubit. Consequently, even a modest 30-logical-qubit program may require close to 10,000 physical qubits—well beyond the scope of near-term quantum hardware.

\vspace{3pt}
\noindent\textbf{Logical T Gate.}
To support universal quantum computation, FTQC circuits are typically decomposed into the Clifford+T gate set~\cite{ross2014optimal}. While Clifford gates are relatively cheap to implement on surface codes~\cite{fowler2018low, bravyi2005universal}, T gates are significantly more costly.
A logical T gate is applied via gate teleportation~\cite{knill2004fault}, using a magic state $|T\rangle = (|0\rangle + e^{i\pi/4}|1\rangle)/\sqrt{2}$. 
%A logical T gate is applied via gate teleportation~\cite{knill2004fault}, using a magic state $|T\rangle = (|0\rangle + e^{i\pi/4}|1\rangle)/\sqrt{2}$ (see Fig.~\ref{fig: FTQCBack}(b)). 
This magic state can be prepared in a surface code patch through \emph{state injection}, which grows a physical magic state into a logical one~\cite{lao2022magic, li2015magic}, but the resulting LER remains at $O(p)$, where $p = 10^{-3} \sim 10^{-4}$, insufficient for fault-tolerant use. 

\begin{figure}[!ht]
\centering
\includegraphics[width=0.42\textwidth]{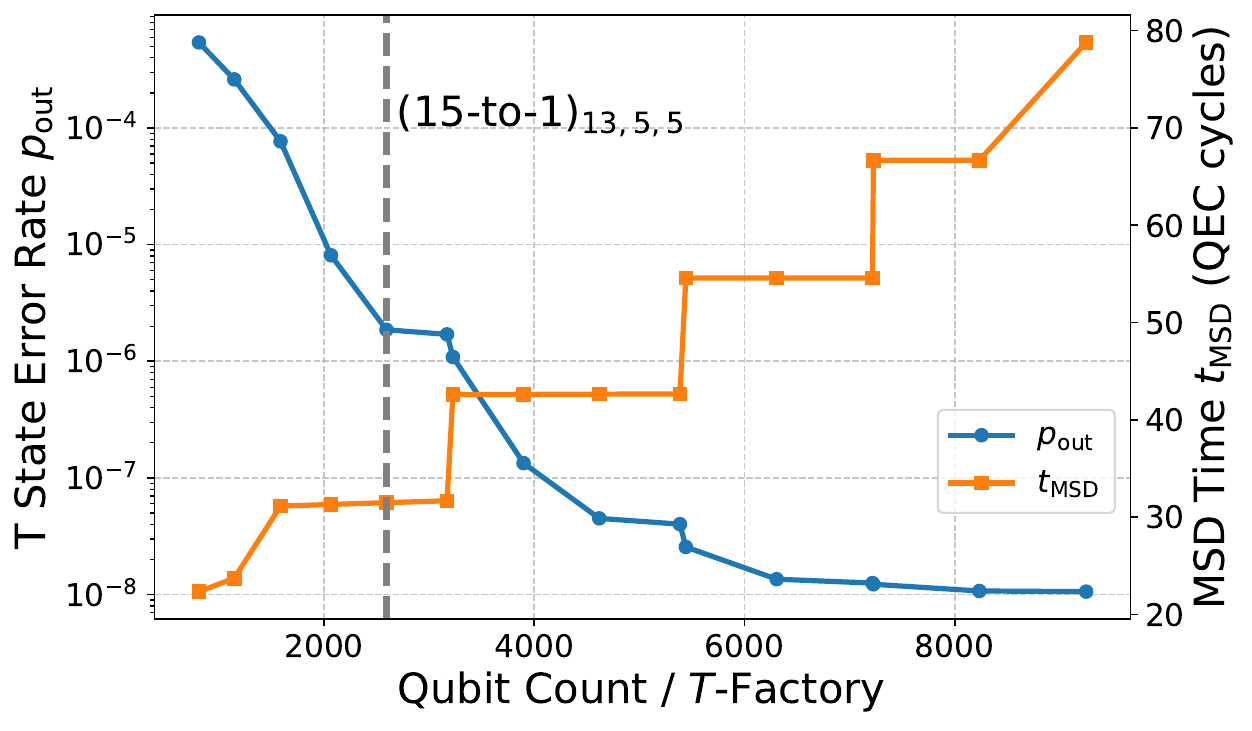}
\caption{State error rate and time cost of different MSD protocols under varying qubit budgets.}
\label{fig: MSD-size}
\end{figure}

%To meet the required LER (e.g., $10^{-6}$), the state usually undergoes magic state distillation (MSD)—a resource-intensive process that consumes multiple noisy magic states to produce one high-fidelity output (Fig.~\ref{fig: FTQCBack}(c)). 
To meet the required LER (e.g., $10^{-6}$), the state usually undergoes magic state distillation (MSD)—a resource-intensive process that consumes multiple noisy magic states to produce one high-fidelity output. 
As shown in \cref{fig: MSD-size}, 
We evaluated the qubit resource and distillation time cost of the $(15\text{-to-}1)_{d_x, d_z, d_m}$ MSD protocol \cite{litinski2019magic}. To meet the distance-9 surface codes's LER, this translates to approximately 2,500 physical qubits for a T gate. Furthermore, approximating a general quantum gate (e.g., $R_z(\theta)$) to precision $10^{-6}$ requires at least $\frac{3}{2}\log_2(10^6) \approx 30$ T gates~\cite{ross2014optimal}, significantly increasing circuit depth and gate count.

While the above FTQC protocol is of foundational theoretical importance and has long been the prevailing approach, recent hardware advances have prompted a shift toward system-level considerations. For the less stringent requirements of EFT, much of the traditional overhead can be circumvented through system-aware design that leverages specific features of trapped-ion hardware and surface codes—as explored in the next section.

\subsection{TIQC Hardware Capability}{\label{subsec: TIQC}}
\label{subsuc: TIQC Background}
Trapped-ion quantum computers (TIQC)~\cite{bruzewicz2019trapped, strohm2024ion} have been a leading quantum platform for over three decades~\cite{cirac1995quantum}, offering unique advantages for realizing EFT.

\vspace{3pt}
\noindent\textbf{Ion Qubits and Gate Operations.}
TIQC systems encode qubits in the internal electronic states of ions confined in linear Paul traps~\cite{paul1990electromagnetic}, where Coulomb repulsion naturally arranges the ions into one-dimensional (1D) chains. These qubits exhibit exceptionally long coherence times—from minutes to days~\cite{bruzewicz2016scalable, strohm2024ion}. Single-qubit (1Q) gates are implemented via tightly focused Raman lasers or global microwave fields, achieving fidelities up to 99.9999\%~\cite{harty2014high}, already meeting the fidelity requirements of EFT. Two-qubit (2Q) gates are realized via the Mølmer–Sørensen (MS) interaction~\cite{molmer1999multiparticle, zhang2025robust, molmer1999multiparticle}, with fidelities reaching 99.97\%~\cite{ballance2014high, srinivas2021high, clark2021high}. Crucially, 2Q gates can be performed between any ion pair within a trap, enabling native all-to-all connectivity. TIQC also supports high-fidelity state preparation and measurement (SPAM) with fidelities exceeding 99.99\%~\cite{myerson2008high}.

\begin{figure}[!ht]
\centering
\includegraphics[width=0.47\textwidth]{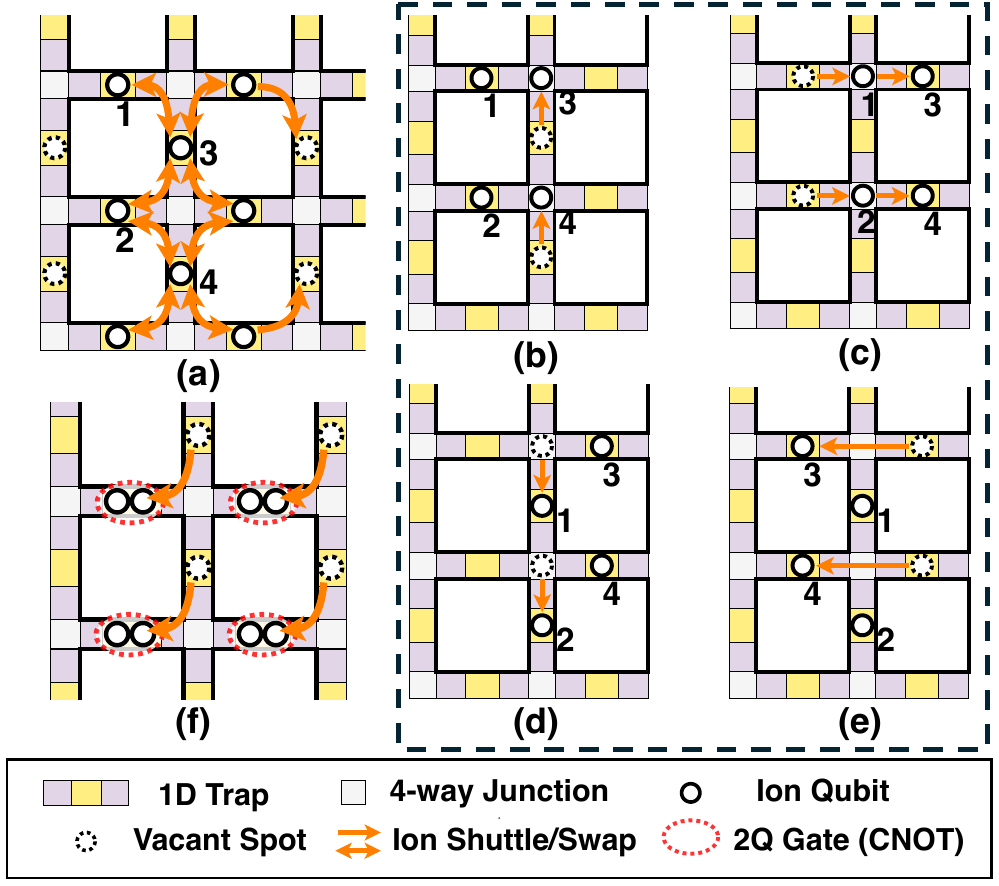}
\caption{(a) QCCD architecture of TIQC, which connects 1D traps with 4-way junctions. It supports ion shuttling and swap. (b-e) The collective ion swap between (1,3) and (2,4) using ion shuttling. (f) 2Q gates (CNOT) are performed on the qubits within the same trap.}
\label{fig: QCCD}
\end{figure}

\vspace{3pt}
\noindent\textbf{QCCD Architecture.}
% While 1D traps provide full connectivity within small ion chains, control becomes increasingly difficult as ion count grows. To scale beyond this limit, the QCCD architecture~\cite{wineland1998experimental, kielpinski2002architecture, blatt2008entangled, murali2020architecting, ovide2024scaling} connects short 1D traps across a 2D grid using junctions~\cite{loschnauer2024scalable, delaney2024scalable}, as illustrated in Fig.~\ref{fig: QCCD}(a). Ions can be shuttled between 1D traps with negligible decoherence~\cite{wineland1998experimental, kielpinski2002architecture}, enabling dynamically reconfigurable connectivity over the grid. For example, Fig.~\ref{fig: QCCD}(b-e) illustrates how the ion shuttling can collectively swap two pairs of ions (1,3) and (2,4). Once two ions are moved into a single trap, a 2Q gate (CNOT) can be performed on them (Fig.~\ref{subsec: TIQC}(f)). This architecture improves scalability and naturally supports QEC codes with 2D layouts, such as surface codes.
While one-dimensional (1D) ion traps provide full connectivity within small chains, scaling them to larger systems introduces significant control complexity. To overcome this, the quantum charge-coupled device (QCCD) architecture~\cite{wineland1998experimental, kielpinski2002architecture, blatt2008entangled, murali2020architecting, ovide2024scaling} connects short 1D traps across a two-dimensional (2D) grid via junctions~\cite{loschnauer2024scalable, delaney2024scalable}, as illustrated in Fig.~\ref{fig: QCCD}(a). 
Ions can be shuttled between 1D traps with negligible decoherence~\cite{wineland1998experimental, kielpinski2002architecture}, enabling dynamically reconfigurable connectivity over the grid. For example, Fig.~\ref{fig: QCCD}(b-e) illustrates how the ion shuttling can collectively swap two pairs of ions (1,3) and (2,4). Once two ions are moved into a single trap, a 2Q gate (CNOT) can be performed (Fig.~\ref{subsec: TIQC}(f)).

% Moreover, the 2D QCCD architecture complements a key characteristic of TIQC platforms: single-qubit gates consistently achieve error rates that are orders of magnitude lower than those of two-qubit gates~\cite{harty2014high, pino2021demonstration, moses2023race}. Moreover, as control electronics and calibration techniques continue to improve, single-qubit gate fidelities are expected to remain high even as the system scales. This motivates the use of QEC to protect the more error-prone two-qubit operations, while preserving the efficiency of high-fidelity single-qubit gates.

\begin{figure*}[!ht]
    \centering
    \includegraphics[width=0.95\textwidth]{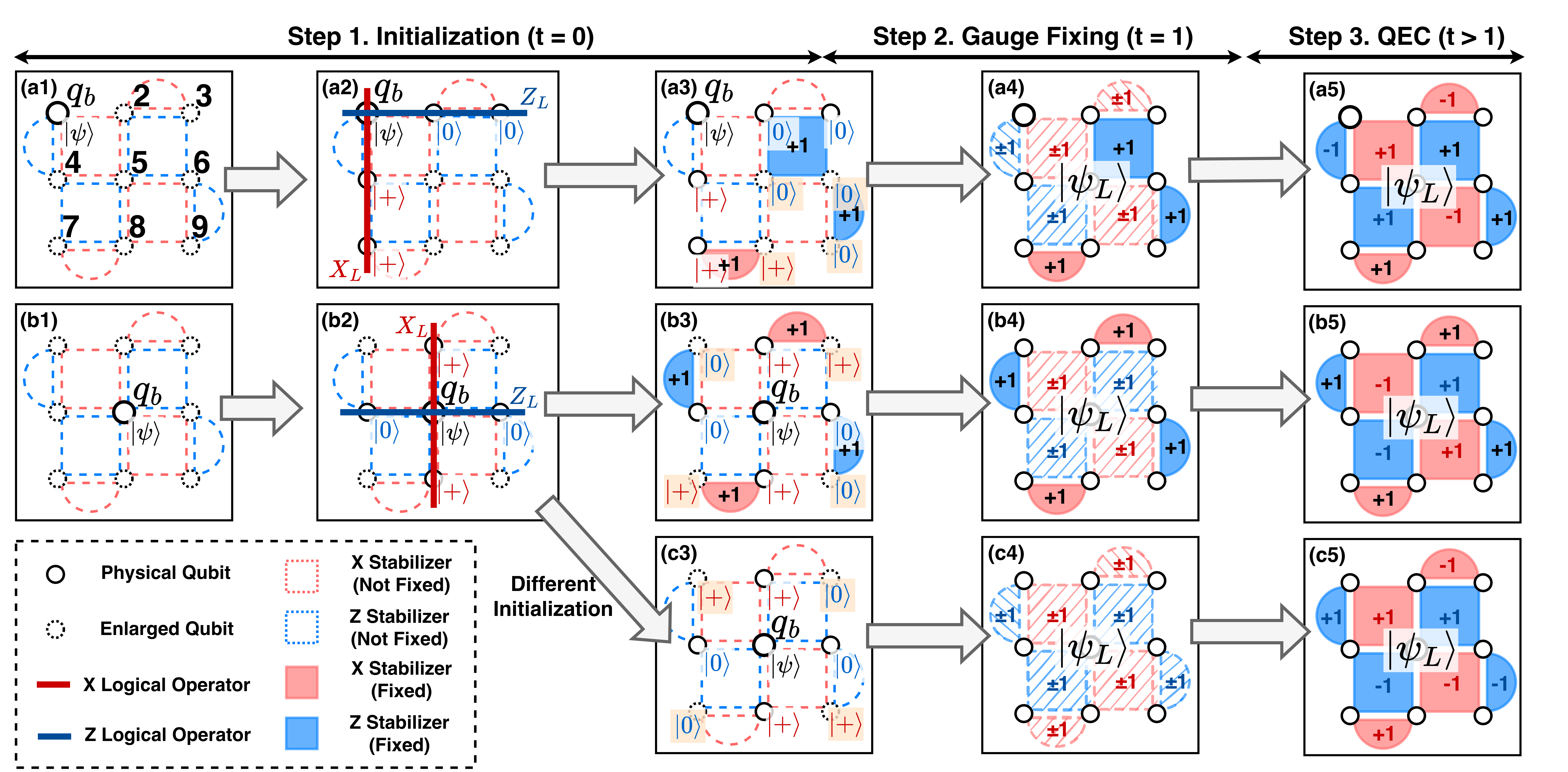}
    \caption{Encoding transition from a bare qubit to a logical qubit under different ancilla initialization configurations. The procedure comprises one round of stabilizer measurement for gauge fixing, followed by $d-1$ repeated rounds for error correction.
    }
    \label{fig:tech1overall}
\end{figure*}

Leveraging these capabilities—ultra-high-fidelity quantum operations, reconfigurable layouts, and architectural scalability—industry leaders such as Quantinuum have published clear roadmaps toward systems with hundreds of qubits~\cite{foss2024progress, quantinuumQuantinuumAccelerates}, positioning TIQC as a leading platform for near-term demonstrations of EFT.

\section{Module I: Runtime Encoding Protocol}
\label{sec:encoding}

This section presents our runtime encoding strategy, inspired by gauge-fixing theory~\cite{vuillot2019code, poulin2005stabilizer}, which converts a bare physical qubit, holding an arbitrary and program-specific quantum state, directly into a fully protected logical qubit encoded in a surface code of any target distance. Crucially, this transformation is performed \emph{in-situ} on the data qubit itself, without the need for preparing additional ancilla logical qubits or post-selecting known ancilla states. Unlike conventional approaches such as state injection~\cite{lao2022magic, li2015magic}, magic state distillation~\cite{bravyi2005universal, bravyi2012magic, haah2018codes}, or FT rotation-based state preparation~\cite{choi2023fault}, our method avoids the extra resource overhead for encoding ancillas and enables direct runtime switching of the data qubit’s encoding.

\subsection{Runtime Encoding Switch}
\label{sec:enlargement-procedure}
This subsection presents the runtime protocol for low-noise, in-situ conversion between bare and logical qubits using surface code patches. Fig.~\ref{fig:tech1overall} shows the encoding switch from a bare to a logical qubit; the reverse process is symmetric and discussed at the end.

% \textcolor{red}{(Jens: would be good if the text in this subsection could refer to (a1), (a2), and the other labels in Fig.~\ref{fig:tech1overall}.  Also, Fig.~\ref{fig:tech1overall} talks about ``Enlarge'', but the text doesn't.  I took a while to realize that what the text calls Step 1, Step 2, and Step 3, actually correspond to the three main segments of Fig.~\ref{fig:tech1overall}.)}

\noindent\textbf{Step 1: Gauge-Aware Ancilla Initialization.} At $t = 0$, we start with a bare qubit $q_b$ in an arbitrary state and prepare surrounding ancilla qubits for logical encoding (Fig.~\ref{fig:tech1overall}(a1, b1)). Ancilla initialization follows the geometry of the target surface code patch and the chosen logical operators $X_L$ and $Z_L$. A natural choice aligns $X_L$ and $Z_L$ along the horizontal and vertical paths intersecting at $q_b$ (Fig.~\ref{fig:tech1overall}(a2, b2)). Ancilla qubits along $X_L$ are initialized in \( \ket{+} \) (stabilized by $X_i$), and those along $Z_L$ in \( \ket{0} \) (stabilized by $Z_i$). Remaining ancilla qubits not on logical operators can be initialized arbitrarily in \( \ket{+} \) or \( \ket{0} \), providing flexibility to reduce gauge operators needing fixing and improve noise control (see Sec.~\ref{Sec: Impact}).

This initialization defines $n-1$ stabilizer generators over $n$ qubits, leaving a single logical degree of freedom at $q_b$, characterized by the intersecting $X$ and $Z$ operators. Gauge-fixing theory ensures that this configuration preserves the bare qubit's logical state during transfer to the surface code.

\noindent\textbf{Step 2: Stabilizer Measurement and Gauge Fixing.} At $t=1$, we perform a round of stabilizer measurements on the target surface code patch, including both the region near the bare qubit and the ancilla-only area. Some stabilizers, already defined by initialization, yield deterministic $+1$ outcomes (solid operators in Fig.~\ref{fig:tech1overall}(a3, b3, c3)). The remaining stabilizers, which anti-commute with those set by initialized qubits, produce random $\pm 1$ outcomes (dashed stabilizers in Fig.~\ref{fig:tech1overall}(a4, b4, c4)). These random outcomes do not compromise subsequent QEC and can be tracked via stabilizer frame updates or corrected with efficient gauge-fixing decoders~\cite{higgott2021subsystem, demarti2024decoding}.

\noindent\textbf{Step 3: QEC Cycles and Code Stabilization.} 
Following gauge fixing, we perform $d-1$ rounds of standard surface-code error correction (Fig.~\ref{fig:tech1overall}(a5, b5, c5)), where $d$ is the target surface code distance. These rounds further suppress noise and allow the decoder to identify and correct residual errors. At the end of this process, the original quantum state \( \ket{\psi} = \alpha\ket{0} + \beta\ket{1} \) is promoted to a fully protected logical state \( \ket{\psi_L} = \alpha\ket{0_L} + \beta\ket{1_L} \), encoded in a surface code patch of distance $d$. The resulting logical qubit is equivalent as one initialized via standard fault-tolerant state preparation.

\noindent\textbf{Reverse Switch: Shrinking Back to Bare Qubit.} Shrinking a logical qubit back to a bare qubit follows the reverse procedure. We measure and discard the ancilla qubits along $X_L$ and $Z_L$ (not including $q_b$), collapsing them in the same basis as their original initialization. For instance, qubits initialized in \( \ket{+} \) are measured in the $X$ basis ($q_4,q_7$ in Fig.~\ref{fig:tech1overall}(a1-a5)). Due to random measurement results in gauge fixing, measuring these ancilla qubits may introduce a global sign flip on the logical operator. For example, if the logical operator is $X_L = X_b X_4 X_7$ (Fig.~\ref{fig:tech1overall}(a1-a5)), and the product of measurement outcomes on $q_4$ and $q_7$ is $-1$, the sign of the $X$ logical operator ($X_b$) has flipped to $-1$. In this case, we apply a correction by flipping the result of the subsequent $X_b$ measurement or applying a $Z_b$ gate. This ensures that the bare qubit accurately reflects the encoded logical state.

% In enlagrement, this is not needed  

% The outcomes are then processed using standard quantum error correction techniques, effectively collapsing the logical patch back into a single unprotected qubit that retains the original logical state $\ket{\phi} = \alpha\ket{0} + \beta\ket{1}$.

\subsection{Impact of Encoding Protocols}
\label{Sec: Impact}
% In the encoding switch, a bare qubit can, in principle, be placed at any data qubit location, as the intersecting row and column at that location naturally define a valid pair of logical $X$ and $Z$ operators. However, different configurations exhibit varying levels of fault-tolerance against circuit-level noise. It is therefore critical to identify an optimal configuration that maximizes the fidelity of the encoding switch.
In the encoding switch, a bare qubit can, in principle, occupy any data qubit, as the intersecting row and column naturally define valid logical $X$ and $Z$ operators. However, the bare qubit's placement and the initialization of surrounding ancilla qubits significantly impact the conversion's logical error rate. Hence, it is critical to identify a configuration that maximizes the fidelity of the encoding switch.

\begin{figure}[!ht]
\centering
\includegraphics[width=0.44\textwidth]{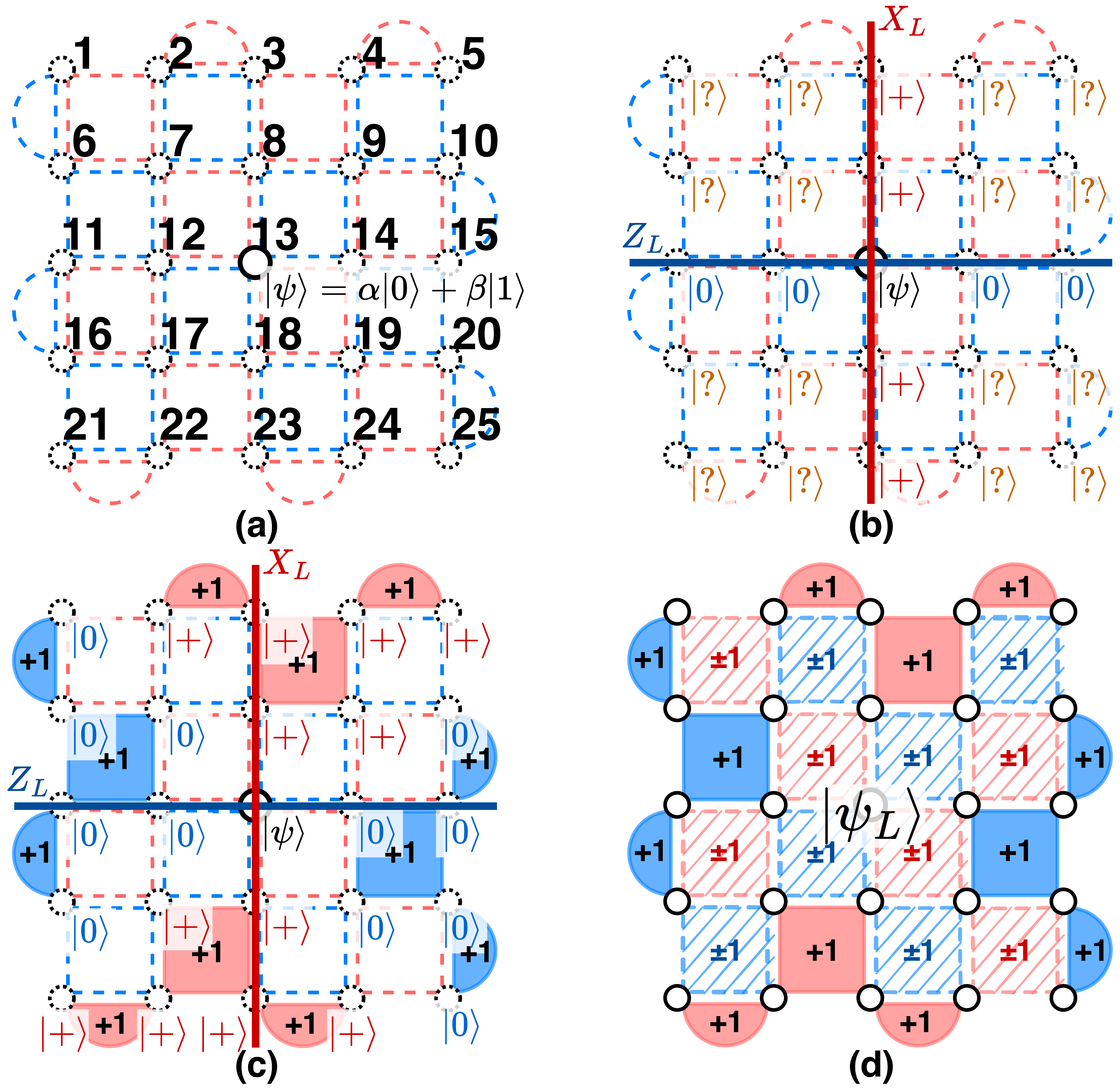}
\caption{Optimized runtime encoding protocol maximizing the number of deterministic stabilizers. 
(a)(b) Bare qubit placement defining logical operators. 
(c) Ancilla initialization pattern with $\ket{+}$ and $\ket{0}$ maximizes deterministic stabilizers. 
(d) Resulting stabilizer measurement outcomes.}
\label{fig: enlarge}
\end{figure}

\begin{table*}[t]
\centering
%\small   
\caption{Hybrid ISA for TIQC: Bare, Logical, and Encoding Switch Instructions}
\label{tab:isa_summary}
\begin{tabular}{p{2cm} p{3.3cm} p{11.2cm}}
\toprule
\textbf{Category} & \textbf{Instruction} & \textbf{Semantics and Purpose} \\
\midrule
\multirow{3}{*}{\textbf{\makecell{Bare Qubit \\ Instructions}}} 
& \texttt{BareMove\_Vertical} & Move a bare ion qubit up/down across rows of trap junctions. \\
%, enabling vertical layout reconfiguration and routing. \\
& \texttt{BareMove\_Horizontal} & Move a bare ion qubit left/right across columns of trap junctions. \\
%within the same trap region or adjacent segments for local connectivity and arrangement. \\
& \texttt{Bare1Q\_Gate} & High-fidelity native single-qubit gates executed on bare qubits. \\% with minimal latency and error. \\
\midrule
\multirow{4}{*}{\textbf{\makecell{Logical Qubit \\ Instructions}}} 
& \texttt{\Movename\_Vertical} & Move a logical patch up/down one step along the trap QEC patch grid. \\%across rows of trap junctions to enable connectivity or co-location. \\
& \texttt{\Movename\_Horizontal} & Move a logical patch left/right one step along the trap QEC patch grid. \\
%Typically used for aligning patches for interaction. \\
& \texttt{\CNOTname\_Transversal} & Perform a logical CNOT gate transversally between two logical on the two overlapped patches.\\
\midrule
\multirow{2}{*}{\textbf{\makecell{Bare–Logical \\ Instructions}}} 
& \texttt{\BLname\_B2L} & Convert a bare qubit into a logical  patch by growing stabilizers region. \\ %Restricted to boundary. \\
& \texttt{\BLname\_L2B} & Convert a logical patch into a bare qubit by shrinking its stabilizer region. \\
%Also restricted to boundary locations. \\
\bottomrule
\end{tabular}
\end{table*}

The key to suppressing errors during the switch lies in preserving as many deterministic stabilizer measurements as possible. Since error correction relies on detecting syndrome changes, stabilizers with random outcomes cannot contribute to identifying errors. For example, in \cref{fig:tech1overall}(a3), a $Z$ error on qubit $q_4$ goes undetected because the error only flips the outcome of the stabilizer $X_{1,2,4,5}$, whose measurement result is already equally likely to be $\pm 1$.

Moreover, whether a stabilizer is deterministic or random depends on the initialization of the ancilla qubits. For instance, in \cref{fig: enlarge}(c), ancilla qubits $q_3$, $q_4$, $q_8$, and $q_9$ are initialized in the \(\ket{+}\) basis. As a result, the operator \(s = X_{3,4,8,9}\) stabilizes the joint state of the bare qubit and the initialized ancillas. During the gauge-fixing step shown in \cref{fig: enlarge}(d), all surface-code stabilizers are measured. Since these stabilizers commute with $s$, its syndrome is expected to remain $+1$, making $s$ a \textit{deterministic stabilizer}.

To maximize conversion fidelity, we introduce a new encoding protocol that initializes ancilla qubits in four triangular regions—two in the \(\ket{+}\) basis and two in the \(\ket{0}\) basis (see \cref{fig: enlarge}(c)). This arrangement ensures that all ancillas are protected by deterministic stabilizers while maintaining balanced region sizes to provide uniform error-correction capability across the layout.

Focusing on $Z$ errors without loss of generality, we find that only $Z$ errors on $q_8$, $q_{13}$, and $q_{18}$ can cause logical faults. For example, a \(Z_3\) error flips the syndromes of both \(X_{2,3}\) and \(X_{3,4,8,9}\), allowing the decoder to detect and correct it unambiguously. In contrast, \(Z_8\) and \(Z_9\) produce identical syndrome patterns—flipping only \(X_{3,4,8,9}\). Our decoder consistently attributes this pattern to \(Z_9\); thus, if the actual error is \(Z_8\), the mis-correction results in a logical $Z$ error during conversion.

Consequently, only a small subset of qubit errors ($Z_8$, $Z_{13}$, $Z_{18}$, $X_{12}$, $X_{13}$, $X_{14}$) can cause logical failure. Importantly, this implies that the probability of conversion-induced logical errors is independent of the surface code distance $d$, meaning that increasing the code distance does not mitigate this error. Therefore, careful hardware-aware architecture design and compiler strategies are essential to efficiently manage and minimize the overhead of encoding-switch operations.

\section{Module II: Hybrid Quantum ISA}
\label{sec:hybrid-isa}

To enable dynamic switching between bare and protected execution, we introduce a hybrid quantum instruction set architecture (ISA) and corresponding microarchitectural designs tailored to TIQC. As illustrated in Fig.~\ref{fig:arch_overview}, the hardware is partitioned into three regions: a \textit{bare qubit region}, a \textit{logical qubit region}, and a \textit{boundary region}. The boundary region serves as a reconfigurable interface between bare and logical domains, where each patch in this region can be interpreted either as a logical surface code block or as a bare qubit before enlarging or after shrinking.
% \textcolor{red}{(Jens: I wonder what ``coarse-grained'' means. This is the only place in the text that talks about ``coarse-grained''.)}

% This architectural flexibility allows operations to be executed at different protection granularities: high-fidelity 1Q gates can run in bare mode, while 2Q gates or critical regions are promoted to FT execution via runtime encoding. The boundary enables seamless bi-directional switching and supports efficient data movement between encoding domains. It also forms the basis for the compiler strategy presented in Sec.~\ref{sec:compiler}, enabling high logical qubit density while avoiding the cost of static encoding, state injection~\cite{lao2022magic}, or magic state distillation~\cite{litinski2019magic}.

\subsection{Hybrid ISA Summary}
\label{sec:isa-summary}
Our hybrid ISA defines three distinct categories of quantum instructions, corresponding to three execution domains: \emph{bare qubit operations}, \emph{logical patch operations}, and \emph{bare–logical switching operations}. This division aligns with our hardware layout and supports efficient compilation and scheduling. 
A summary of the hybrid ISA is provided in Table~\ref{tab:isa_summary}.

\noindent \textbf{Bare Qubit Instructions.} These instructions operate on unencoded qubits in the bare region. Designed to leverage the native strengths of TIQC, they enable high-fidelity operations with minimal overhead. \texttt{BareMove} perform ion swaps across traps to dynamically reconfigure bare qubit layouts. \texttt{Bare1Q\_Gate} allows arbitrary native rotations to be directly applied to bare qubits, avoiding the need for QEC overhead when protection is not required.

\noindent \textbf{Logical Patch Instructions.}
These instructions operate on encoded qubits within surface code patches. To enable flexible layout and interactions, logical qubits must be movable while preserving their stabilizer structure. The \texttt{\Movename} instructions allow movement across the 2D QCCD grid, supporting logical gate reconfiguration without requiring teleportation or decoding.

\noindent \textbf{Bare–Logical Switching Instructions.} To support hybrid execution, the ISA includes two dedicated instructions — \texttt{\BLname\_B2L} and \texttt{\BLname\_L2B} — for runtime conversion between encoding domains. These operations are restricted to the boundary region, which simplifies scheduling and localizes the coordination between bare and logical regions. This constraint ensures modularity and minimizes inter-domain coupling.

\noindent \textbf{Design Rationale.}
The ISA is deliberately minimal and modular, reducing architectural complexity while maintaining flexibility. By isolating switching operations to explicit boundary instructions, the compiler can optimize independently within each domain, ensuring scalable and error-aware scheduling. The dual-granularity design, which uses bare qubits for high-fidelity, low-overhead operations, and logical qubits for protected execution, enables tunable trade-offs between fidelity, resource usage, and performance, laying a robust foundation for dynamic compilation (Sec.~\ref{sec:compiler}).

\begin{figure}[!ht]
\centering
\includegraphics[width=0.47\textwidth]{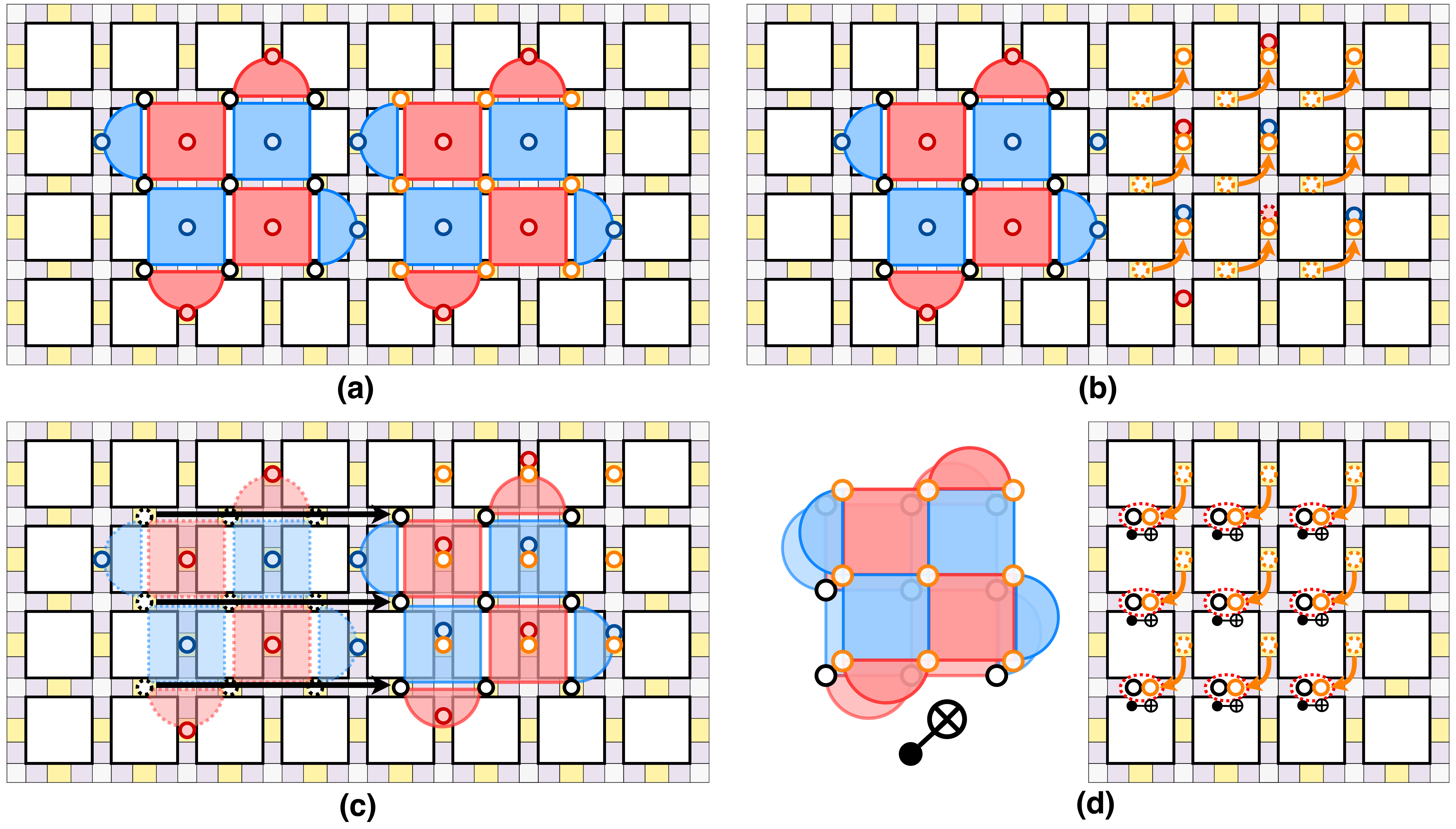}
\caption{(a) Mapping surface-code patches onto a 2D QCCD grid.
(b) Patches are moved to clear routing channels.
(c) Patches are aligned into an overlapping region for interaction.
(d) Return the patch to perform transversal CNOT.}
\label{fig: logCNOT}
\end{figure}

\subsection{Microarchitecture Designs}
Our architectural design targets 2D junction-based ion trap systems (Sec.~\ref{subsec: TIQC}), inspired by recent Quantinuum hardware and surface-code-based ion trap schemes proposed in~\cite{loschnauer2024scalable, delaney2024scalable, leblond2023tiscc}. 
We divide the hardware into bare and logical regions, and ensure hybrid ISA can be implemented efficiently across both domains.
We divide the hardware into bare and logical regions, and ensure instructions in Table~\ref{tab:isa_summary} can be implemented efficiently across both domains.

\noindent \textbf{\texttt{\Movename.}} 
Fig.~\ref{fig: logCNOT}(a) illustrates the implementation of the surface code in our architecture, while Fig.~\ref{fig: logCNOT}(b--c) shows the fine-grained steps involved in logical patch movement.
The highlighted yellow regions denote computation zones that support either two-qubit physical gates or single-qubit ion measurements.
Each one-dimensional (1D) trap can host multiple physical qubits; for clarity of exposition, we initially place a single data qubit in each 1D trap.
Ancilla ions are placed along vertical junctions and serve as syndrome qubits for stabilizer measurements.
Movement of adjacent logical patches is realized via ion shuttling across trap junctions.

\noindent \textbf{\texttt{\CNOTname\_Transversal.}}
To implement transversal CNOT gates between two logical patches, we leverage 1D traps that can contain multiple ions. As shown in Fig.~\ref{fig: logCNOT}(d), logical patches are moved into proximity such that each 1D trap hosts a pair of qubits—one from each patch. CNOT gates are then applied in parallel across all pairs. After execution, the patches are separated by reversing the movement steps. This approach ensures FT logical CNOT without requiring teleportation or measurement-based gates.

\noindent \textbf{\texttt{\BLname}.} 
At the boundary region, \texttt{\BLname\_B2L} and \texttt{\BLname\_L2B} instructions are invoked to transition data between unencoded and protected forms (Fig.~\ref{fig: encoding}). These operations restrict deformation to boundary region, which significantly simplifies both architectural design and scheduling. As a result, intra-region operations can remain fully independent, avoiding compiler-level entanglement between instruction types.

\begin{figure}[!ht]
\centering
\includegraphics[width=0.45\textwidth]{Fig/Compile.pdf}
\caption{Switch between bare ($P_1$) and logical qubits ($Q_{2,3}$).}
\label{fig: encoding}
\end{figure}

%\section{\frameworkname~Module III: Compiler with Dynamic Encoding}
\section{Module III: Compiler Support}
\label{sec:compiler}
Building on the hybrid ISA and region-aware architecture introduced in Sec.~\ref{sec:hybrid-isa}, we develop a fidelity- and resource-aware compiler that manages both the encoding state and spatial placement of program qubits. It addresses two main challenges: (1) \textit{dynamic encoding allocation}, deciding when and where qubits reside in logical versus bare form, and (2) \textit{routing and co-location of logical patches} to enable two-qubit gates efficiently.

When the logical patch region can encode all program qubits, the compiler can perform \texttt{CodeSwitch} in-place, removing the need for a separate physical qubit region.  "In-place" means the data qubit does not need to move to the boundary. When ancilla resources are sufficient, the ancillas released after an \texttt{L2B} operation for logical qubit $Q_{1}$ can remain locally associated with $Q_{1}$ and be reused in its next \texttt{B2L}, instead of being reassigned to another qubit $Q_{2}$.
A qubit can temporarily switch to bare form for a single-qubit $R_z$ gate and immediately switch back to a logical patch for protection, avoiding any physical movement or routing overhead.

In practice, however, the number of active surface code patches is limited, constraining the “logical register” space, while the bare region provides flexible, lower-fidelity memory. Additionally, grid-based adjacency restricts gate execution, often requiring explicit qubit movement.

To handle these constraints, the compiler leverages abstractions inspired by classical register allocation and NISQ-era routing. The resulting compilation passes generate low-overhead, error-aware schedules that respect encoding constraints while balancing fidelity, locality, and resource usage.

\subsection{Bare-Logical Encoding Allocation}
A central challenge in compiling programs for \frameworkname~is dynamically deciding which qubits should reside in logical patches and which can remain as bare qubits, analogous to classical register allocation with limited fast-access registers.

In \frameworkname, logical patches function as a finite register file, while bare qubits serve as main memory. Program operations fall into two categories: (1) high-fidelity single-qubit non-Clifford gates on bare qubits, and (2) two-qubit gates requiring both qubits in logical form. Minimizing \emph{spills}—runtime conversions that incur overhead and introduce additional error—is the primary compilation goal. We model encoding allocation as a temporal coloring problem over the program’s qubit usage graph and employ a greedy linear-scan algorithm inspired by classical register allocation.

At each timestep, a two-qubit gate triggers logical encoding for both operands; if capacity is insufficient, idle logical qubits are shrunk to bare form to free space. Single-qubit gates are executed on bare qubits, postponing conversion until necessary. This temporal reuse strategy reduces conversions and maximizes fidelity.

The compiler maintains a pool of logical qubits limited by the available logical region. Conversions are applied only when required, and logical qubits are proactively shrunk once protection is no longer needed, freeing space for subsequent allocations. This mirrors classical register allocation and spilling, while incorporating a fidelity-aware model in which each conversion introduces additional noise.

\subsection{Routing: Connectivity Management}
Beyond encoding allocation, the compiler must also ensure that logical qubits involved in 2Q operations are physically adjacent on the ion trap grid. To this end, we adopt a SABRE-inspired routing strategy~\cite{li2019tackling}, treating logical patches as movable units and scheduling swaps to co-locate interacting qubits. Given the 2D layout of the logical patch region, we construct a logical coupling graph and apply SABRE to compute low-cost swap schedules. Each proposed \texttt{Swap} operation is translated into our hybrid ISA using \texttt{\Movename\_Vertical} or \texttt{\Movename\_Horizontal} instructions, depending on the alignment of the operand patches.

A similar routing strategy is applied within the bare region, where qubit movement is implemented via physical ion swaps. At the interface between bare and logical zones, we incorporate conversion constraints into the routing logic to jointly optimize conversion and movement steps. This co-optimization enables efficient spatial scheduling while maintaining correctness and minimizing decoherence.

\section{Evaluation}\label{sec: eval}

\subsection{Experiment Setup}
\label{sec: eval setup}
\noindent \textbf{Baselines.}  
We compare \frameworkname~against three baselines:

\noindent (1) \textbf{\NISQname:} A NISQ-only baseline in which all operations are executed on bare physical qubits without quantum error correction. This approach incurs minimal qubit overhead but is highly vulnerable to noise.

\noindent (2) \textbf{\MSDname:} A fully fault-tolerant baseline that uses surface code with distance-$d$ encoding for all program qubits. All gates are implemented fault-tolerantly. Non-Clifford $R_Z(\theta)$ gates are decomposed into $T$ gates using Gridsynth~\cite{ross2014optimal}, and each $T$ gate is executed via a distilled magic state prepared using the $(15\text{-to-}1)_{13,5,5}$ protocol~\cite{litinski2019magic}. As a result, \MSDname~must allocate a significant portion of its qubit budget to magic state distillation (MSD), thereby reducing the number of qubits available for encoding program logical qubits.

\noindent (3) \textbf{\INJname:} A partially fault-tolerant baseline that implements non-Clifford $R_Z(\theta)$ gates via state injection~\cite{dangwal2025variational, lao2022magic}. The protocol first prepares the $R_Z(\theta)$ state on a small $d=3$ code patch using post-selection, then enlarges the patch to form an ancilla logical qubit compatible with the program qubits. The logical $R_Z(\theta)$ gate is then applied via teleportation using logical CNOT gates and measurements.

\vspace{3pt}
\noindent\textbf{Noise Model.}  
To simulate realistic trapped-ion quantum computing (TIQC) behavior, we use a gate-level noise model calibrated with recent experimental benchmarks~\cite{moses2023race, pino2021demonstration, harty2014high, bruzewicz2019trapped}. The error rates are derived from our randomized benchmarking experiments on the Quantinuum’s H1 system, as shown in \cref{fig: error break}(a), with single-qubit gates at \(1 \times 10^{-5}\), two-qubit gates at \(10^{-3}\), SPAM (state preparation and measurement) errors at \(10^{-3}\), idling errors at \(10^{-5}\) and shuttling errors at \(10^{-4}\), consistent with current TIQC performance~\cite{quantinuum-h1, quantinuum-h2}.

\vspace{3pt}
\noindent \textbf{Benchmarks.} We evaluate our approach on variational quantum algorithms using the \textit{unitary coupled-cluster singles and doubles} (UCCSD) ansatz~\cite{sarkar2022modular}, a standard choice for VQAs. In particular, our benchmark design follows recent EFT studies~\cite{dangwal2025variational}, ensuring comparability with state-injection-based evaluation. We focus on ground-state energy estimation tasks from the physics models and quantum chemistry \cite{singh2020ising, bonechi1992heisenberg}, relevant to EFT quantum computing.

\iffalse
\noindent (1) \textit{Physics models}, including the Ising~\cite{singh2020ising} and Heisenberg~\cite{bonechi1992heisenberg} Hamiltonians. These models are widely used to benchmark near-term quantum devices due to their simple structure and fundamental relevance to condensed matter physics. Their qubit count can be flexibly scaled with system size; we evaluate instances with $n = 10, 20, 30$ qubits to reflect increasing problem complexity while remaining within the scope of EFT-capable systems.

\noindent (2) \textit{Quantum chemistry}, focusing on ground-state energy estimation of small molecules such as H$_2$O and BeH$_2$~\cite{vqe_qiskit_repo}. Unlike the physics models, these problems require encoding molecular electronic structures into qubits via fermion-to-qubit transformations, resulting in a fixed qubit count ($n=10$) determined by the number of spin orbitals. 
\fi

% Additionally, we investigate the Hamiltonians of the $\text{H}_2\text{O}$ and $\text{BeH}_2$ molecules. The molecular configurations are as follows:
% \begin{itemize}
%     \item $\text{H}_2\text{O}$ - $ \ell = 0.96\,\text{\AA}(\times2) $ - $ ^1\text{A}_1$
%     \item $\text{BeH}_2$ - $ \ell = 0.69\text{\AA}(\times2)$ - $ ^1\Sigma^+$
% \end{itemize}

\vspace{3pt}
\noindent \textbf{Metrics.}  
We adopt two primary metrics for end-to-end performance evaluation: (1) \textit{VQA energy} $E_s$, which denotes the final converged energy produced by the VQA circuit. As proposed in~\cite{mcclean2016theory}, lower energy values correspond to better performance. To quantify improvement, we compute the ratio of the energy gap between the baseline and the ideal (noiseless) energy to that achieved by \frameworkname. A higher ratio indicates greater performance gain. (2) \textit{Program fidelity} $f$, defined as the probability that a benchmark completes without any logical error. Higher fidelity implies better reliability and correctness. To quantify the improvement of \frameworkname~over baseline A, we define the fidelity improvement ratio as  
$
\gamma = \frac{1 - f_{\text{A}}}{1 - f_{\frameworkname}}.
$
A larger value of $\gamma$ indicates that \frameworkname~achieves a greater reduction of infidelity.

\vspace{3pt}
\noindent \textbf{Evaluation Setting.}  
To balance simulation accuracy and efficiency, we follow simulation strategies from previous works~\cite{dangwal2025variational, lao2022magic} for small- and large-scale circuits. Evaluation was conducted on a server with an AMD EPYC 9534 64-core processor and 1.48 TB of memory.

\vspace{3pt}
\noindent \textit{Small-scale circuits ($\leq$10 qubits).}  
For small circuits, we use the Qiskit-Aer simulator~\cite{Qiskit-Aer} in density matrix mode to model noise based on realistic hardware parameters. 
% VQA circuit optimization is done using the Nelder–Mead algorithm~\cite{nelder1965simplex}.

\vspace{3pt}
\noindent \textit{Large-scale circuits ($>$10 qubits).}  
% For larger circuits, we use a scalable approximation by restricting the ansatz to Clifford circuits, which efficiently approximate low-energy behavior \cite{ravi2022cafqa, seifert2024clapton}. 
Discrete parameter optimization is performed with a classical genetic algorithm~\cite{wright1991genetic, holland1992genetic} using tournament selection, crossover, and mutation over 50 generations with a population size of 64.

\newlength{\figheight}
\setlength{\figheight}{3.35cm}
\begin{figure*}[!ht]
    \centering
     (a) Noise Model Validation \hspace{0.7cm} (b) \CNOTname\_Transversal \hspace{0.7cm} (c) \Movename\_Vertical \hspace{0.8cm} (d) \BLname\_B2L\\
    \includegraphics[height=\figheight]{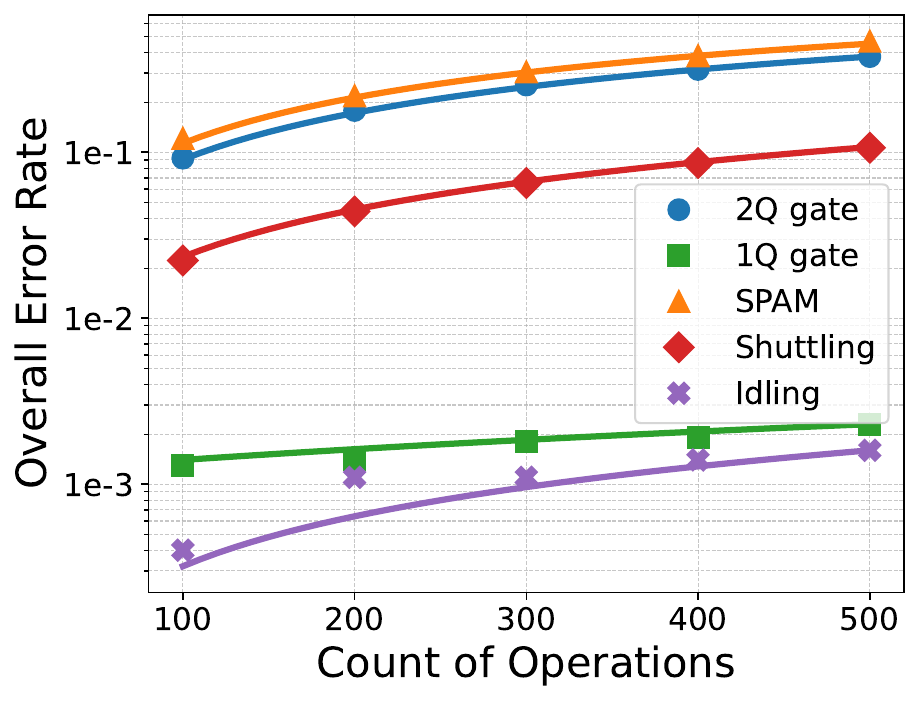}
    \includegraphics[height=\figheight]{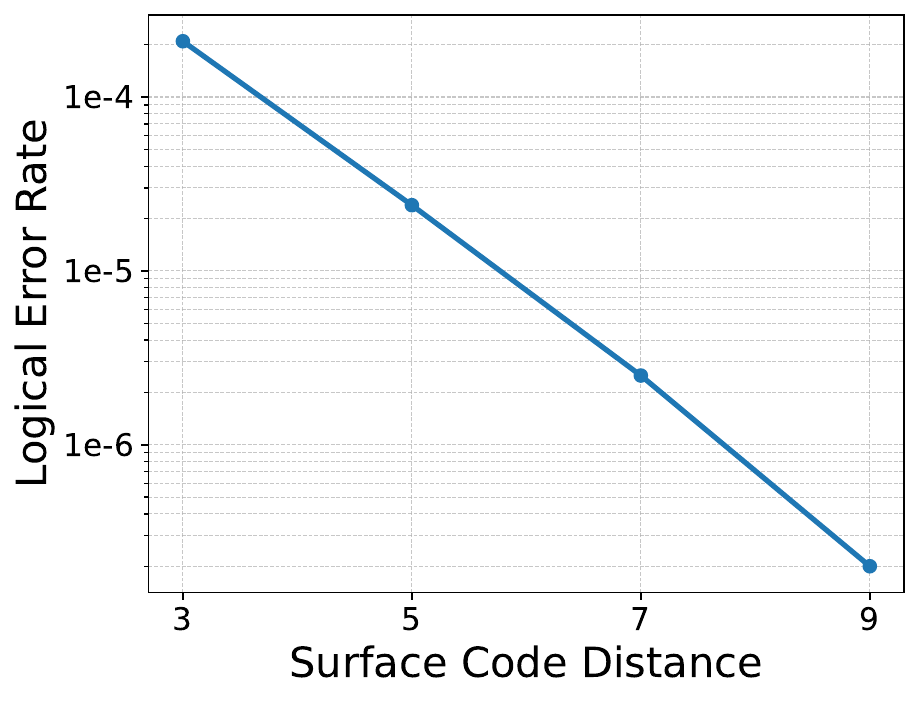}
    \includegraphics[height=\figheight]{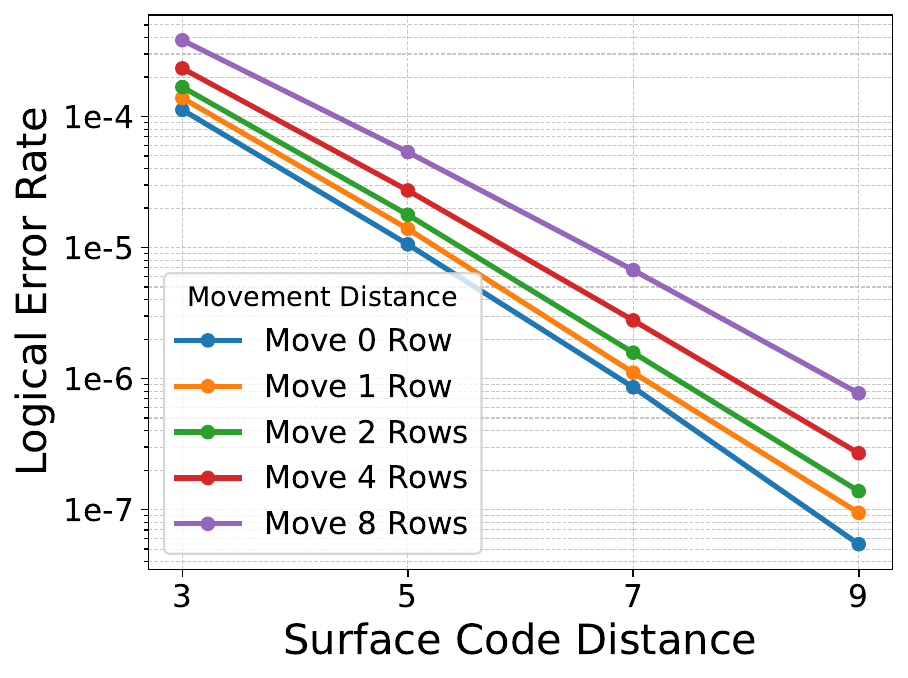}
    \includegraphics[height=\figheight]{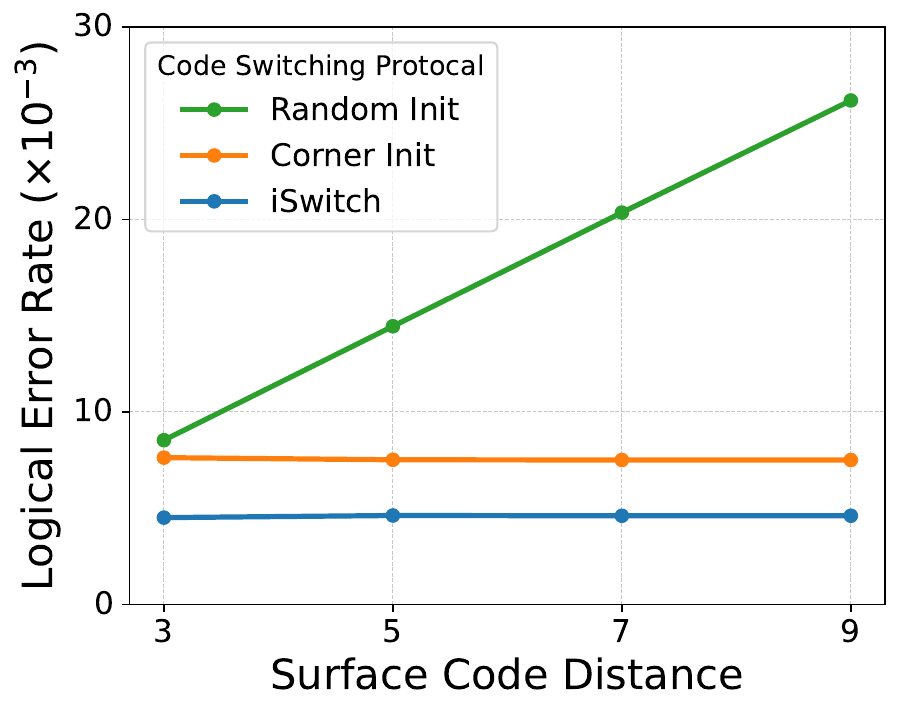}
    \caption{(a) Experiment results on a Quantinuum's H1 device to measure noise channel parameters. (b)–(d) Logical error rate simulations using the experimentally calibrated noise model, showing the logical error rates for (b) transversal 2-qubit gates, (c) logical patch movement, and (d) code conversion operations across varying code distances.}
    \label{fig: error break}
\end{figure*}

\subsection{Simulation of Logical Operations}
We first validate our noise model on the Quantinuum H1 device~\cite{quantinuum-h1}, and then use the validated model to simulate the LERs of the operations (\CNOTname, \Movename, \BLname) defined in \cref{tab:isa_summary} across various code distances.

\noindent \textbf{1. Noise model validation.} 
In Fig.~\ref{fig: error break}(a), we compare the overall error rate accumulation as a function of operation count for different physical qubit operations, including CNOT errors, SPAM errors, idling errors (with and without qubit shuttling), and single-qubit gate errors. The close agreement between experimental \textit{markers} and simulation \textit{lines} confirms the accuracy of our model, providing a solid foundation for subsequent logical operation simulations.

\noindent \textbf{2. Logical error rate simulation.} 
In Fig.~\ref{fig: error break}(b--d) We simulate the LERs of three logical operations with Stim~\cite{stim} and Pymatching~\cite{higgott2022pymatching}. Our key observations are as follows: 

\noindent \textit{(i)} The LERs of \CNOTname~and \Movename~decrease with increasing code distance, demonstrating fault-tolerance. In contrast, \BLname~remains independent of code distance, reflecting the partial fault-tolerance inherent in our \frameworkname.

\noindent \textit{(ii)} In Fig.~\ref{fig: error break}(c), the LER of patch movement increases with the number of rows or columns traversed. Here, "move 1" corresponds to moving a patch to overlap with an adjacent patch, while "move 0" indicates that the patch remains idle. This trend arises from the accumulation of idling errors in the absence of intermediate QEC. However, a single distance-2 move still introduces fewer errors than two consecutive distance-1 moves, which can be attributed to the extra patch movement required to clear the path and return (\cref{fig: logCNOT}(b)(d)).

\noindent \textit{(iii) Comparison of runtime encoding strategies (Fig.~\ref{fig: error break}d).} We consider three protocols: (1) \textit{\frameworkname}, (2) \textit{Corner Enlargement}, which performs the switch at the patch corner, and (3) \textit{Random Initial State}, which targets the patch center but randomly initializes ancilla qubits in $|0\rangle$ or $|+\rangle$ states during enlargement (see Sec.~\ref{sec:enlargement-procedure}). Our in-center-switch strategy achieves the lowest LER by maximizing the number of deterministic stabilizers around the injected state, thereby protecting the conversion process. In contrast, random initialization performs progressively worse for larger codes, as the absence of deterministic stabilizers allows nearly any qubit error to propagate into a logical error, and the probability of such errors increases with code size.

\subsection{Comparison to NISQ Approach}
\begin{figure*}[!ht]
    \centering
    \includegraphics[width=0.98\textwidth]{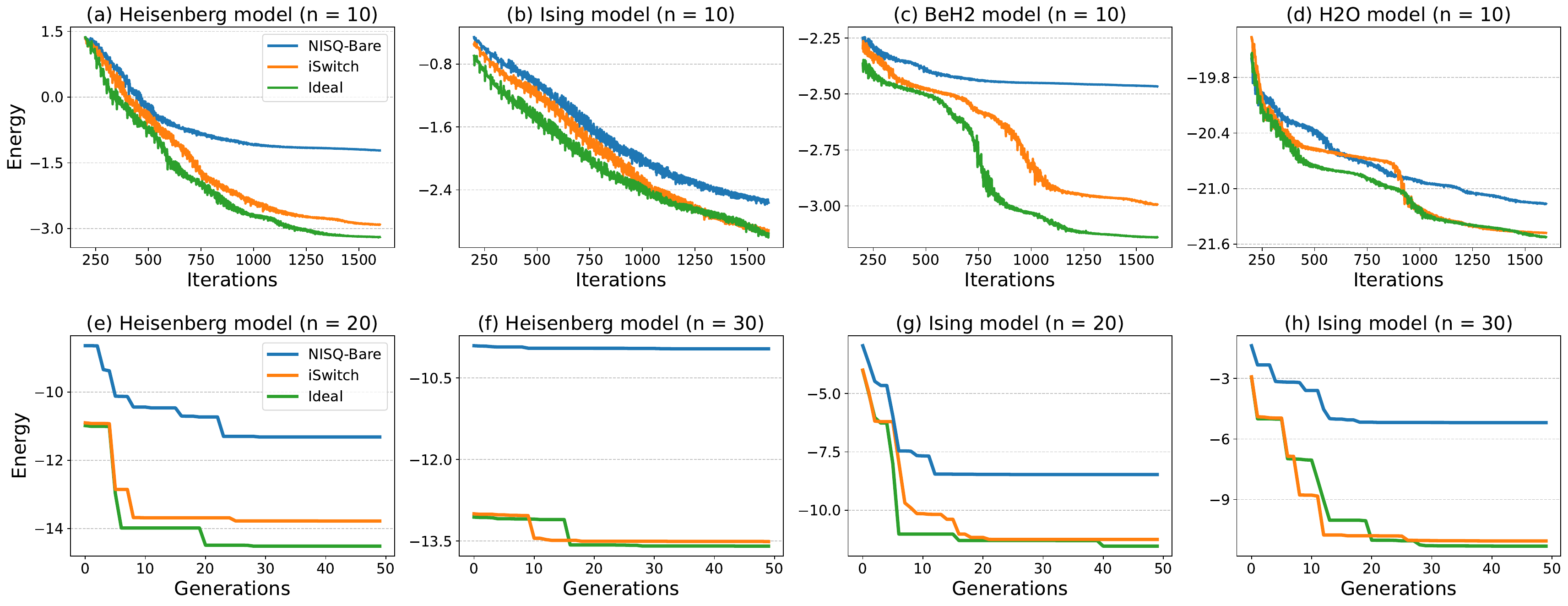}
    \caption{VQA energy comparison of \frameworkname~($d=9$) and \NISQname. Ideal denotes the noiseless VQA circuit.}
    \label{fig:iSwitch vs NISQ}
\end{figure*}

We evaluate the improvement of \frameworkname, which uses a surface code with \(d=9\), over \NISQname~by comparing the final converged VQA energies, as shown in \cref{fig:iSwitch vs NISQ}, which reflect end-to-end performance. An "Ideal" reference line represents the noiseless circuit performance. To quantify improvement, we define the energy improvement ratio as the gap between \NISQname~and the ideal baseline, divided by the gap under \frameworkname. \frameworkname~consistently achieves lower VQA energies. For small-scale cases (\cref{fig:iSwitch vs NISQ}(a)–(d)), the improvement ratios are 6.89×, 13.1×, 4.64×, and 7.38×, respectively. For large-scale benchmarks (\cref{fig:iSwitch vs NISQ}(e)–(h)), the gains increase to 4.34×, 43.4×, 10.8×, and 23.6×.  

This improvement arises from three key factors.  
First, in \frameworkname, all 2Q gates are QEC-protected with LERs below \(10^{-6}\), whereas in \NISQname, they remain unprotected with LERs near \(10^{-3}\). This difference in fidelity significantly reduces accumulated noise. 
Second, in VQA circuits, the number of 2Q gates is much larger than the number of non-Clifford gates, making them the primary source of error. As the circuit size increases, the gap between the number of 2Q gates and non-Clifford gates grows, leading to a greater reduction in noise for larger circuits in \frameworkname. 
Third, our compiler minimizes noisy transitions between bare and logical qubits, trading a small number of code-switching operations to protect all 2Q gates on logical qubits, thus avoiding noisy 2Q operations on bare qubits.

% Fig.~\ref{fig: eval-three}(a1,a2) examines the impact of surface code distance in \frameworkname. Beyond $d=9$, both VQA energy and fidelity saturate, as logical CNOT errors are already suppressed, and conversion fidelity becomes the dominant bottleneck. In contrast, small code distances (e.g., $d=3$ or $d=5$) offer insufficient protection—at $d=3$, performance even degrades compared to \NISQname~due to syndrome extraction overhead and conversion noise.  
% At the optimal setting of $d=9$, \frameworkname~achieves 3.48×, 4.52×, 3.54×, and 4.68× reductions in infidelity across the four large-scale benchmarks.

% Third, VQA circuits are inherently robust to small amounts of local noise. With an approximate $4\times$ reduction in infidelity, the residual errors become sufficiently sparse and localized for the circuit to tolerate, resulting in a $20$--$40\times$ improvement in the ground-state energy gap relative to the ideal case.

\subsection{Comparison to QEC Approaches}
In this section, we evaluate the improvement of \frameworkname~over two current QEC methods: fully-QEC \MSDname~and partial-QEC \INJname. In \cref{fig: Comp to Logic}, we compare the program fidelity of \frameworkname~(orange), \INJname~(purple), and \MSDname~(red) across four benchmarks.

\noindent \textbf{Comparison to MSD-Based Method.}  
As shown in \cref{fig: Comp to Logic}, \MSDname~requires, on average, $2.06\times$ more qubits to achieve the same fidelity as \frameworkname. While \MSDname~can achieve higher fidelity with sufficient qubit resources (as it is a fully fault-tolerant method), \frameworkname~remains superior for larger programs (n > 30) with equivalent qubit resources.

This improvement stems from two main factors:
First, \MSDname~allocates a significant portion of its qubit budget to magic state factories, forcing the use of lower-distance encoding for program qubits. This results in higher LER of 2Q gates due to insufficient protection. For an example, with a budget of 4860 qubits, \MSDname~uses 2594 qubits to construct the T-factory, leaving only enough qubits to encode program qubits using distance-6 surface codes. 

Second, \MSDname~requires decomposing all non-Clifford single-qubit gates (e.g., \(R_Z(\theta)\)) into Clifford+T gates, leading to significant circuit depth overhead. The increased depth exacerbates noise accumulation, as logical qubits must idle while a target qubit sequentially executes multiple \(T\) gates.

\vspace{3pt}
\noindent \textbf{Comparison to Injection-Based Method.} As shown in \cref{fig: Comp to Logic}, \INJname~requires, on average, $1.49\times$ more qubits to achieve the same fidelity as \frameworkname. Both methods reach a fidelity ceiling, as they are partial fault-tolerant, and overall program fidelity is constrained by the non-FT \(R_Z\) gates.  
The advantages of \frameworkname~arise from the following factors:

First, \INJname~requires all program qubits to be fully encoded and additionally relies on approximately 50\% more logical qubits~\cite{dangwal2025variational} for ancilla qubits to prepare and store injected \(R_Z(\theta)\) states. This approach reduces flexibility of encoding ratio compared to \frameworkname's hybrid design and requires extra resources for implementing the \(R_Z\) gate.

Second, injection-based implementations of \(R_Z(\theta)\) often require multiple ancillary states, such as \(R_Z(\theta)\), \(R_Z(2\theta)\), and \(R_Z(4\theta)\), to probabilistically construct a single \(R_Z(\theta)\) rotation. The state injection protocol itself is probabilistic, and there is a 50\% chance that an injected \(R_Z(\theta)\) state will result in \(R_Z(-\theta)\), requiring the injection of an additional \(R_Z(2\theta)\) state. This probabilistic nature means that, on average, two injected \(R_Z\) states are needed to construct a single \(R_Z(\theta)\) gate, with the number of states to prepare being uncertain in advance.

Third, the success probability of these state injection protocols decreases as the surface code distance increases. As a result, injection methods often initialize the state on a small code (e.g., \(d=3\)) and gradually enlarge it, which increases circuit depth and exposure to noise. Additionally, because post-selection is used in preparing the \(R_Z(\theta)\) states, the probability of successfully generating the state decreases with increasing code distance. This gradual enlargement of the surface code increases circuit depth, and during this process, the generated \(R_Z\) state is more susceptible to noise due to insufficient code distance.

\vspace{3pt}
\noindent \textbf{Scalability analysis.}
We observe an empirical \emph{diminishing-return point} beyond which the fully fault-tolerant (\MSDname) execution begins to match or exceed the performance of \frameworkname, such that allocating additional physical qubits no longer yields meaningful fidelity gains for EFT.
For the evaluated $20$ logical-qubit workloads, this transition occurs around $\sim 5$k physical qubits (\cref{fig: Comp to Logic} (a)(c)).

Importantly, this transition is \emph{not} a fixed qubit-scale threshold, but shifts with the \emph{program scale}.
For $30$ logical-qubit workloads, the diminishing point moves to around $\sim 7$k physical qubits. The performance gap is governed by the effective physical-qubit budget per logical qubit. For a fixed device size, increasing the number of logical qubits reduces this per-logical budget and therefore delays the regime in which Full-FT overtakes EFT.

\subsection{Component Justification Analysis.}
\noindent\textbf{Impact of Surface Code Distance.}  
We study how varying the surface code distance affects the performance of \frameworkname, as measured by the fidelity improvement ratio over \NISQname~(\cref{fig: ablation study}(a)).
(1) The results indicate that increasing the code distance beyond \(d=9\) yields only marginal improvements in program fidelity. This is becasure Beyond \(d=9\), the logical CNOT error becomes sufficiently suppressed, and conversion fidelity becomes the dominant performance bottleneck, which further increasing the code distance cannot mitigate.  
(2) In contrast, small-distance surface codes fail to provide adequate protection. At \(d=5\), the fidelity improvement is minimal, and at \(d=3\), performance is even worse than the \NISQname~baseline—resulting in higher VQA energy—due to the overhead of syndrome measurements and conversion-related noise.

\setlength{\figheight}{3.65cm}
\begin{figure*}[!ht]
    \centering
     (a) Heisenberg (n = 20) \hspace{1.1cm} (b) Heisenberg (n = 30) \hspace{1.5cm} (c) Ising (n = 20) \hspace{2.0cm} (d) Ising (n = 30)  \hspace{0.4cm} \\
    \hspace{-0.3cm} \includegraphics[height=\figheight]{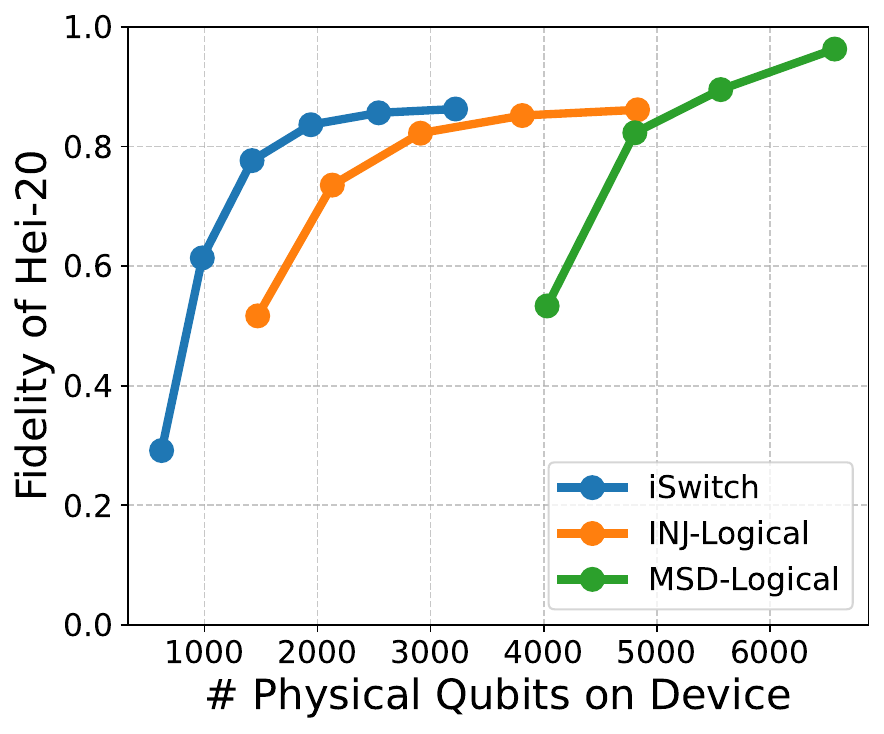}
    \includegraphics[height=\figheight]{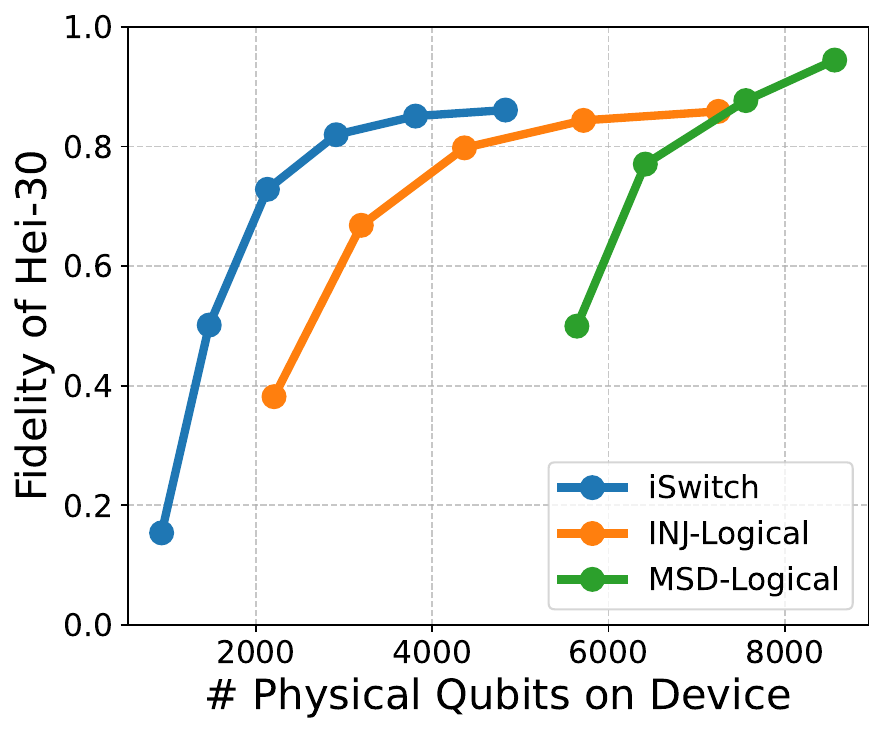}
    \includegraphics[height=\figheight]{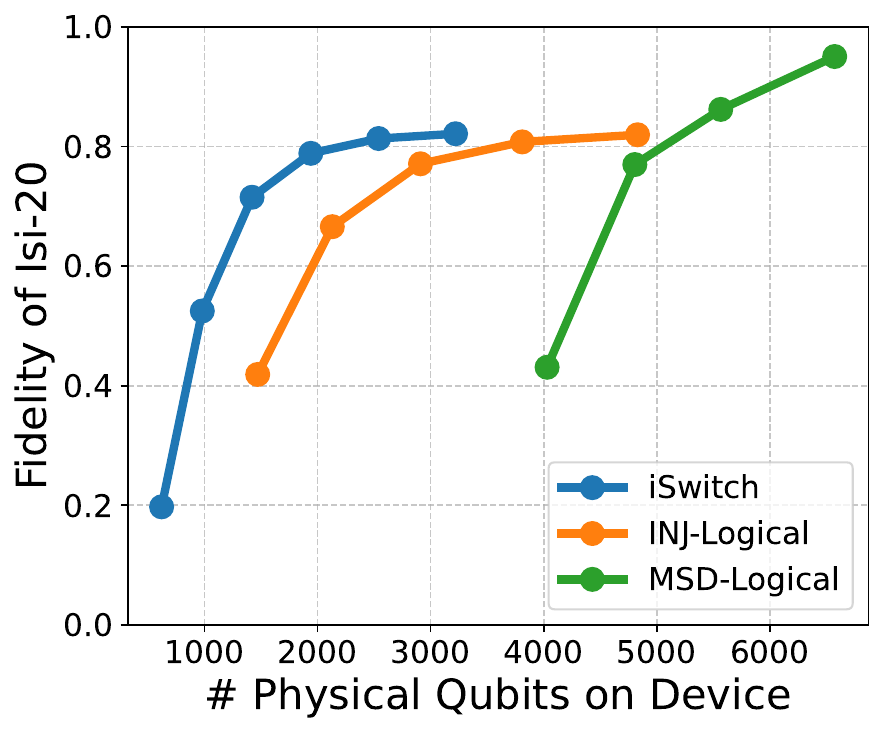}
    \includegraphics[height=\figheight]{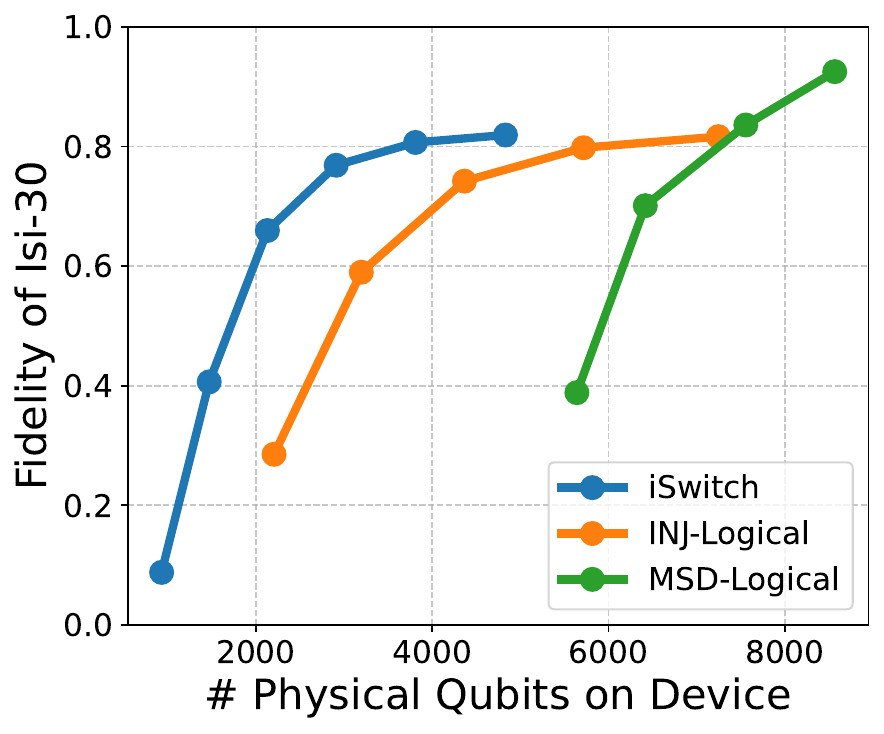}
    \caption{Program fidelity comparison between \frameworkname~(orange) and other QEC approaches, including state injection (purple) and magic state distillation (red), across varying device sizes and benchmarks.}
    % \Description{Intro.}
    \label{fig: Comp to Logic}
\end{figure*}

\setlength{\figheight}{3.5cm}
\begin{figure*}[!ht]
    \centering
     (a) \hspace{5.25cm} (b) \hspace{5.25cm} (c) \\
    \includegraphics[height=\figheight]{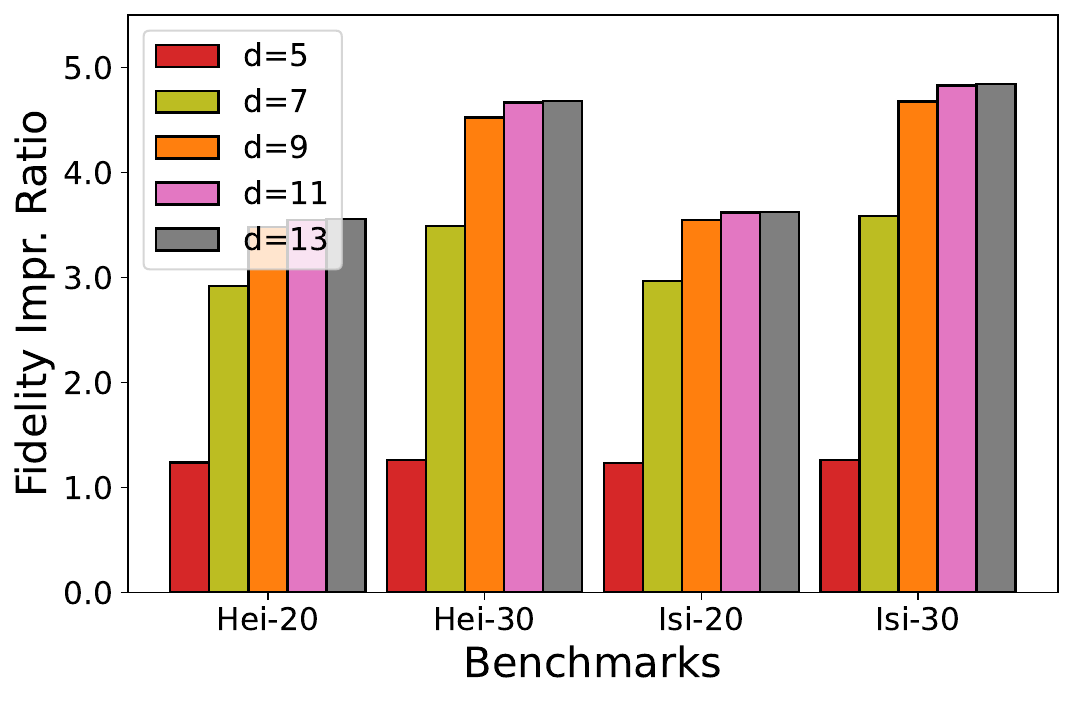}
    \hspace{0.3cm}
    \includegraphics[height=\figheight]{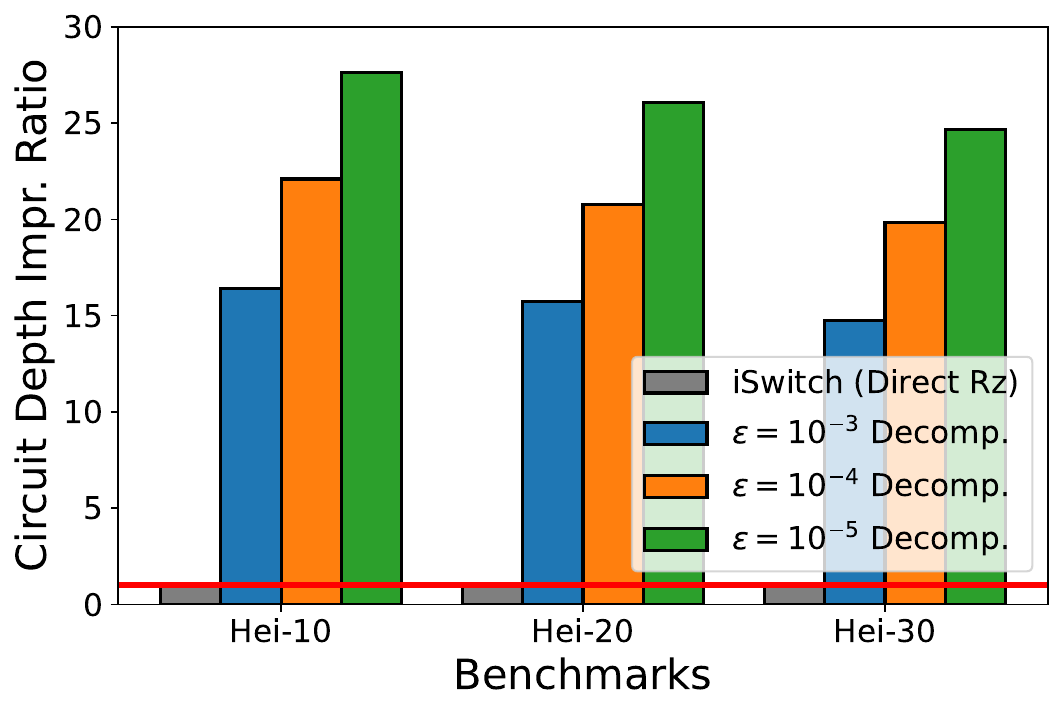}
    \hspace{0.3cm}
    \includegraphics[height=\figheight]{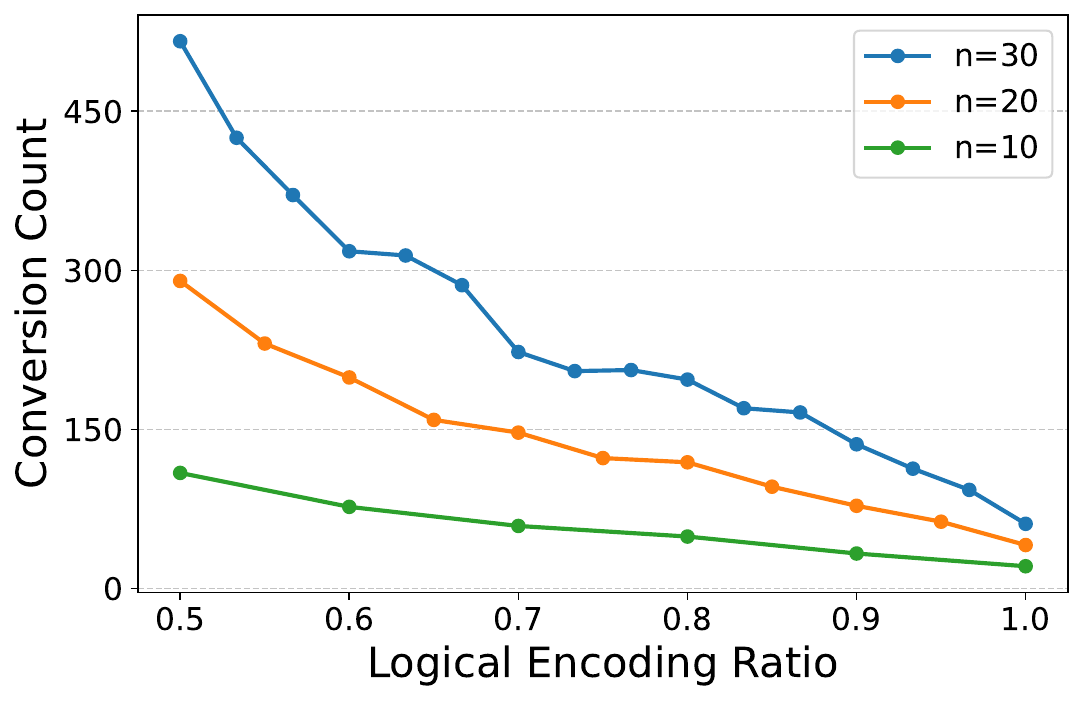}
    \caption{Ablation study of \frameworkname. (a) Impact of varying code distance on program fidelity. (b) Impact of omitting Clifford+T decomposition on circuit depth reduction. (c) Impact of varying encoding ratio on the number of encoding conversions.}
    % \Description{Intro.}
    \label{fig: ablation study}
\end{figure*}

\vspace{3pt}
\noindent\textbf{Impact of omitting Clifford+T Decomposition.}  
In \frameworkname, non-Clifford single-qubit gates (e.g., \(R_Z(\theta)\)) are executed directly on bare qubits, avoiding Clifford+T decomposition. This significantly reduces circuit depth: as shown in \cref{fig: ablation study}(b), conventional MSD-based methods incur 14.9×–27.8× depth increases depending on decomposition accuracy \(\epsilon\). The shorter circuits lower qubit idling errors, reducing noise accumulation from waiting for T-state distillation. 
Furthermore, because each \(R_Z(\theta)\) gate is decomposed into multiple T gates, maintaining low accumulated error requires high-fidelity T states. This, in turn, demands a high-quality MSD factory, which can consume substantial qubit resources.

\vspace{3pt}
\noindent\textbf{Impact of Logical Encoding Ratio.}
We evaluate how limiting the number of simultaneously active logical qubits affects performance. Specifically, we vary the \textit{logical encoding ratio}—defined as the maximum number of concurrently encoded logical qubits divided by the total number of program qubits—and measure its impact on runtime behavior. As shown in \cref{fig: ablation study}(c), reducing the logical encoding ratio increases the number of encoding-switch operations required. This is expected: with fewer logical qubits allowed, qubits must more frequently transition between bare and logical forms to satisfy gate dependencies, increasing swap and routing activity across architectural regions.

\section{Conclusion}

We introduced \frameworkname, a compiler-assisted selective QEC framework for TIQC. By in-situ encoding switching, it executes single-qubit non-Clifford gates with zero extra ancilla logical overhead and maintains high-fidelity two-qubit operations, reducing qubit and conversion cost compared to fully fault-tolerant and existing early fault-tolerant methods. These results demonstrate practical, near-term fault tolerance on hardware-constrained trapped-ion devices.

\section{Acknowledgments}
We thank the anonymous reviewers for their constructive feedback, and Ang Li from PNNL for helpful discussions.
This material is based upon work supported by the U.S. Department of Energy, Office of Science, National Quantum Information Science Research Centers, Quantum Science Center (QSC). 
This work was supported in part by NSF Grants 2048144, NSF 2422169, NSF 2427109, NSF 2519029 and PHY-2110421, as well as ARO Grant W911NF-20-1-0037.
Additional support is provided by NSF National Quantum Virtual Laboratory (NQVL) under NSF 2435382 and the NSF Challenge Institute for Quantum Computation (CIQC) under NSF 2016245.
This work was also supported in part by the Cisco Research Fund and benefited from NVIDIA CUDA-Q and Quantinuum Systems hardware promotional offers.

% ---------- References ----------
\bibliographystyle{unsrt}
\bibliography{references}

\end{document}